\documentclass[11pt]{article}

\DeclareUnicodeCharacter{03B2}{$\beta$}
\DeclareUnicodeCharacter{03BC}{$\mu$}
\DeclareUnicodeCharacter{2212}{-}
\DeclareUnicodeCharacter{2500}{-}
\DeclareUnicodeCharacter{03BD}{$\nu$}
\DeclareUnicodeCharacter{03C4}{$\tau$}
\DeclareUnicodeCharacter{03C5}{$\upsilon$}
\DeclareUnicodeCharacter{03B1}{$\alpha$}

\usepackage[a4paper,margin=1in]{geometry}
\usepackage{amsmath}
\usepackage{amssymb}
\usepackage{graphicx}
\usepackage{xspace}
\usepackage{xcolor}
\usepackage{lineno}
\usepackage[numbers,super,sort&compress]{natbib}
\usepackage{setspace}
\usepackage{url}

\newcommand{\bp}{$\beta$~Pic\xspace}
\providecommand{\gaia}{{Gaia}\,DR3}

\graphicspath{{figures/}}

\title{\vspace{-2em}\textbf{Discovery of radio emission from the exoplanet $\beta$~Pictoris~b}}
\author{K.~N.~Ortiz Ceballos$^{1\ast}$, E.~Berger$^{1}$, Y.~Cendes$^{2}$ \\[0.3em]
\small $^{1}$Center for Astrophysics $\mid$ Harvard \& Smithsonian, 60 Garden St, Cambridge, MA 02138, USA\\
\small $^{2}$Department of Physics and Astronomy, University of Oregon, Eugene, OR 97403, USA\\
\small $^{\ast}$e-mail: kortizceballos@cfa.harvard.edu}
\date{}

\begin{document}
\maketitle
\doublespacing

\noindent
\textbf{Planetary magnetic fields shape atmospheric escape, mediate interactions with stellar winds, and encode information about planetary interiors\cite{brain_exoplanet_2024}, yet they have not been directly measured for planets beyond the Solar System. A direct observable signature is auroral radio emission produced by the electron cyclotron maser instability, whose highest emitted frequency is set by the magnetic field strength at its source\cite{treumann_electron-cyclotron_2006}. Although auroral radio bursts are observed in Solar System planets and in some ultracool dwarfs\cite{burke_observations_1955,berger_discovery_2001,kao_strongest_2018}, no radio detection has previously been unambiguously localized to an extrasolar planet rather than its host star\cite{vedantham_coherent_2020,callingham_radio_2024}. Here, we report the first direct detection of auroral radio emission from an exoplanet, the 
giant planet $\beta$~Pictoris~b, with the MeerKAT array. We detect rapid, recurring, and highly circularly polarized bursts, as well as persistent emission, at frequencies of 0.85 to 3.5 GHz. We identify the emission as electron cyclotron maser radiation, which implies a magnetic field of $\gtrsim 1.25$ kG at the planet --- the first such direct field strength measurement for an exoplanet.
}

\vspace{1em}

Auroral radio emission is a direct, model-independent probe of a planet's magnetic field; it consists of coherent emission from electrons accelerated along magnetic field lines, and it bears signatures of planetary rotation, magnetic obliquity and magnetospheric plasma content\cite{hess_modeling_2011,dulk_complete_1992}. In the Solar System, Earth and the four giant planets all emit such radiation, at kilometric (Earth, Saturn, Uranus, Neptune) and decametric (Jupiter) wavelengths\cite{zarka_auroral_1998}.
Similarly, gigahertz-frequency, strongly-polarized bursts have been detected\cite{berger_discovery_2001,hallinan_periodic_2007,berger_periodic_2009} in ultracool dwarfs (UCDs) down to the planetary mass regime\cite{kao_strongest_2018}, which are phenomenologically similar to planetary aurorae\cite{hallinan_magnetospherically_2015,williams_variable_2017}.
Massive exoplanets thus offer natural targets for radio emission searches, but a coherent radio signal can also arise from a magnetically active host star\cite{villadsen_ultra-wideband_2019,callingham_population_2021}, and previous searches of directly-imaged massive exoplanets yielded only upper limits\cite{cendes_pilot_2022}, including at megahertz frequencies for \bp~b itself\cite{shiohira_search_2024}, or, where coherent emission was found in a planet-hosting system, it could not be spatially assigned to the planet rather than the star\cite{vedantham_coherent_2020,turner_search_2021,zhang_circularly_2025,turner_tentative_2026}. 

$\beta$~Pictoris (HD~39060) is an A6V star located at a distance\cite{gaia_collaboration_gaia_2023} of $19.63\pm0.06$ pc in the $\sim 23$ Myr old $\beta$~Pictoris moving group\cite{lee_revisiting_2024}, hosting a debris disk\cite{smith_circumstellar_1984} and at least three giant planets\cite{lagrange_probable_2009,lagrange_giant_2010,lagrange_evidence_2019,gibbs_discovery_2026,sutlieff_direct_2026}. The most massive planet in the system, \bp~b (mass, $M\approx 12$ M$_{\rm Jup}$; semi-major axis, $a\approx 10$ AU)\cite{lacour_mass_2021}, reaches an angular separation of up to $\approx 0.55''$ across its 24-year orbit, and the host star is magnetically quiet\cite{zwintz_revisiting_2019}, making the system an ideal target for radio observations. 
With a spectral type of L2$\pm$1 and $T_{\rm eff}\approx1700$K\cite{chilcote_1-24_2017}, \bp~b is a young, planetary-mass analog of the UCDs in which auroral emission has been detected: it matches the L0+L1.5 binary 2MASSW J0746425+200032 in spectral type\cite{berger_periodic_2009,zhang_multiepoch_2020} and the T2.5 dwarf SIMP J013656.5+093347 in mass\cite{kao_strongest_2018} ($\sim$13~$M_{\rm Jup}$), but with a dynamically-confirmed planetary mass and formation rather than a field brown dwarf origin. We observed the \bp system with the MeerKAT array on four occasions in 2025 and 2026, in the L~band (0.8--1.7~GHz) and S~band (1.7--3.5~GHz); see Extended Data Table~1; Methods.

\begin{figure}[t]\centering
\includegraphics[width=\linewidth]{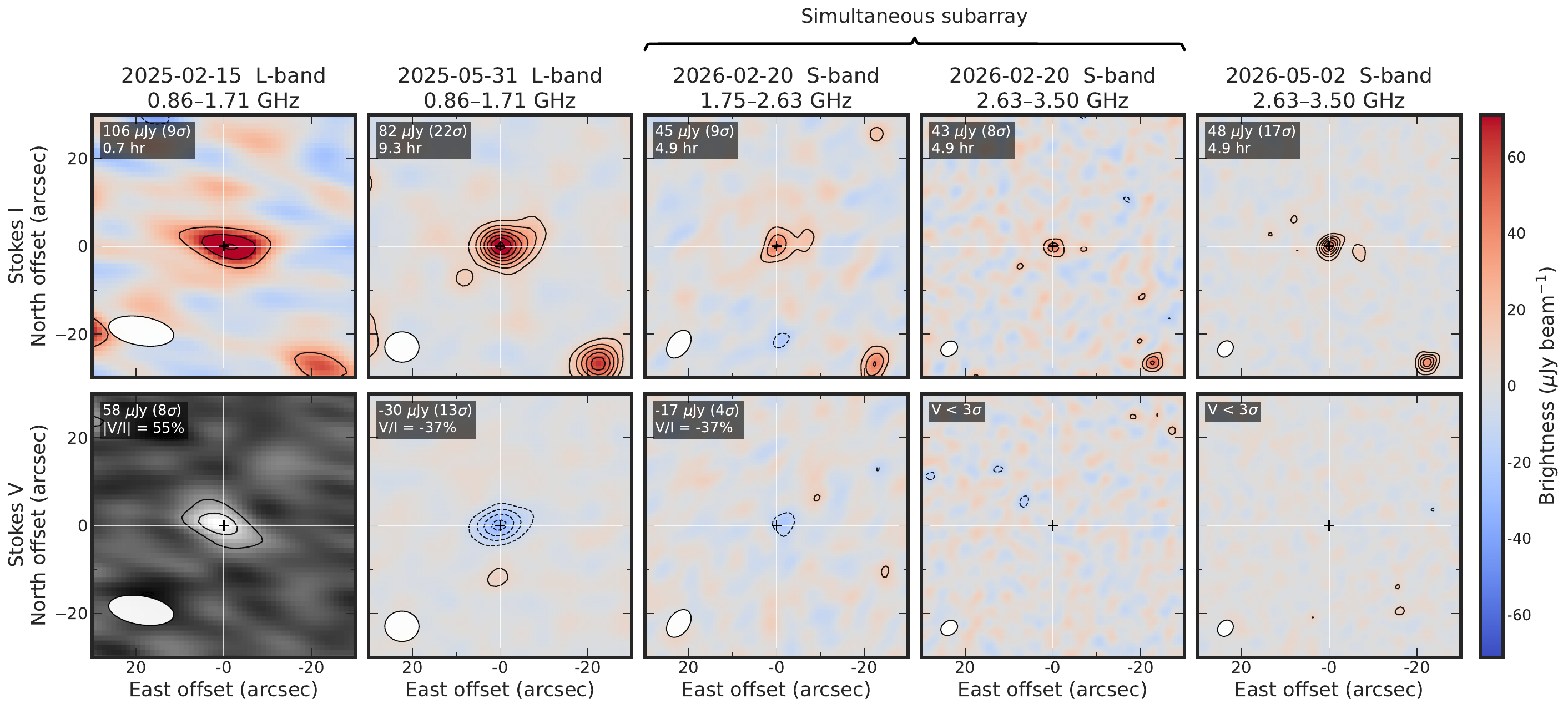}
\caption{\textbf{Radio detections of \bp~b in four observing sessions and at two frequency bands.} Shown are full-track integrations of the total intensity (Stokes~$I$, top) and circularly polarized (Stokes $V$, bottom) images centered on the position of \bp~b after proper- and orbital-motion propagation. The contours indicate signal-to-noise ratios ranging from 3 to 21, in steps of 3 (dashed contours are negative flux), while red and blue colors indicate positive and negative flux densities, respectively. Negative flux densities in Stokes $V$ indicate left-handed circular polarization; in the first observation the handedness of polarization is not constrained and we use a greyscale colormap instead (Methods).}
\label{fig:cutouts}
\end{figure}

\begin{figure}[t]\centering
\includegraphics[width=0.8\linewidth]{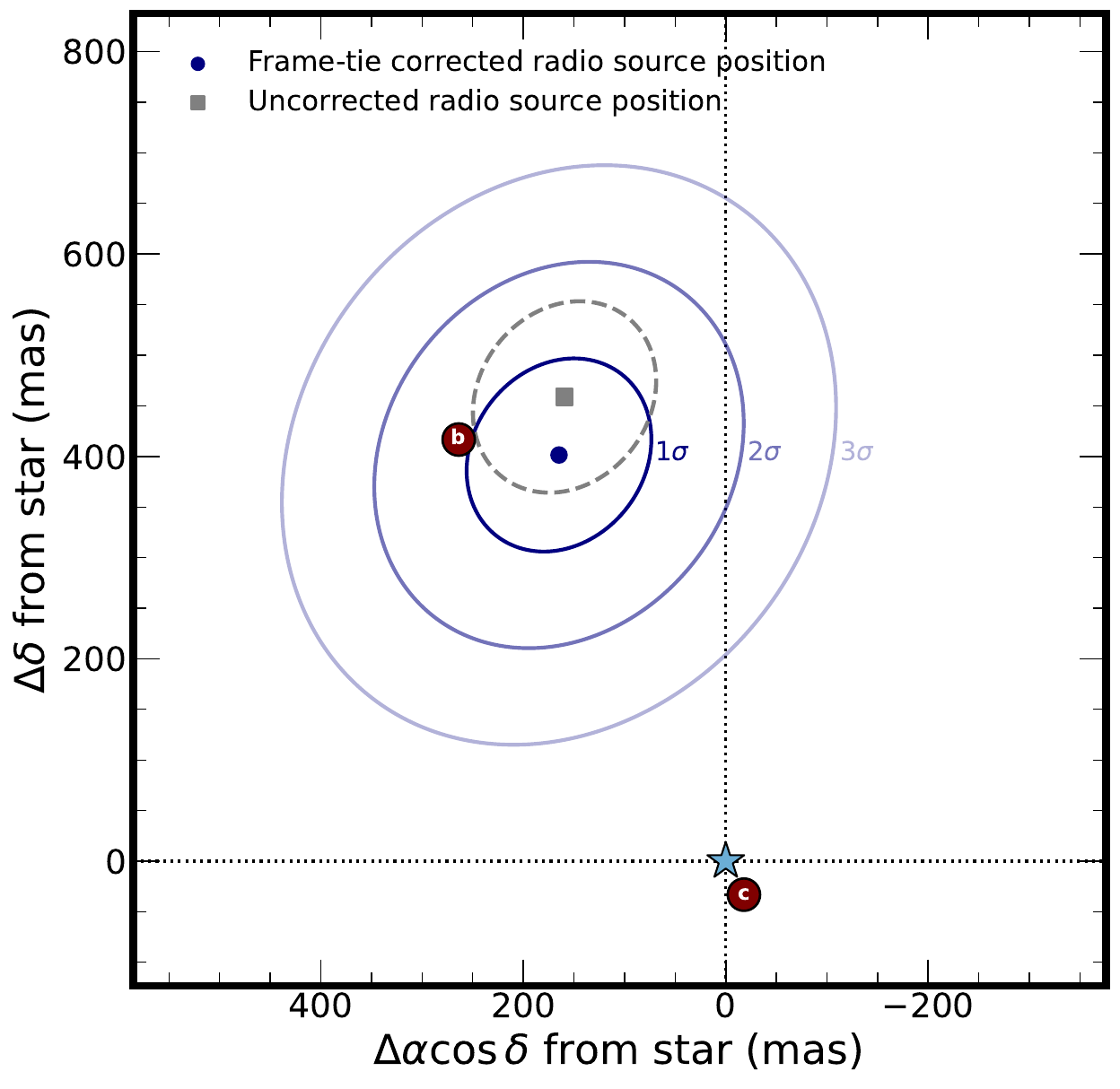}
\caption{\textbf{Localization of the radio emission to \bp~b.} Frame-tie-corrected radio position (blue; 1,2,3$\sigma$ covariance ellipses from the Monte Carlo error propagation, including the systematics inflation; Methods) relative to the star $\beta$~Pictoris (blue star) and its planets b and c (red circles). The uncorrected radio source position is shown for reference in grey, and is consistent with the corrected position within 1 standard deviation. The radio source coincides with $\beta$ Pic b (Mahalanobis radius $R=1.1$, $p=0.53$) and is significantly inconsistent with the host star ($R=4.9$, $4.4\sigma$) and planet~c ($R=5.3$,
$4.8\sigma$) (Methods). The outermost planet in the system, planet d, lies outside of the frame.}
\label{fig:astrometry}
\end{figure}

In all epochs, we detect a radio source at the location of the \bp system (Fig.~\ref{fig:cutouts}; Methods). 
The decisive question is whether the emission originates at one of the planets or the host star. 
We tie the radio image to the {\it Gaia} celestial reference frame using compact radio counterparts of {9} {\it Gaia}-identified quasars detected across the field, supplemented by one VLBI calibrator, and solve for an affine frame-tie correction\cite{driessen_21_2022,driessen_frb_2024} (translation plus rotation, scale and shear; Methods, Extended Data Table~{2}). After applying the correction (which shifts the position by less than its $1\sigma$ uncertainty) and propagating all uncertainties by Monte Carlo sampling, the radio source coincides with the known position of \bp~b (Fig.~\ref{fig:astrometry}; $p\approx 0.53$) and is highly inconsistent with the host star at $4.4\sigma$ significance ($p\approx 6\times10^{-6}$) and with planet~c at $4.8\sigma$ significance ($p\approx 8\times10^{-7}$), accounting for all measured systematics ($5.2\sigma$ and $5.6\sigma$, respectively, with statistical uncertainties alone). The incorporated systematics are based on two empirical tests (Methods): a frequency-differential test of ionospheric refraction, which finds a median per-sightline chromatic residual of only 32~mas (Extended Data Fig.~2), and a point source injection recovery test, which shows the fitted position uncertainty to be accurate to $\approx 1.1$ of the fitting covariance (Extended Data Fig.~5).

\begin{figure}[t]\centering
\includegraphics[width=\linewidth]{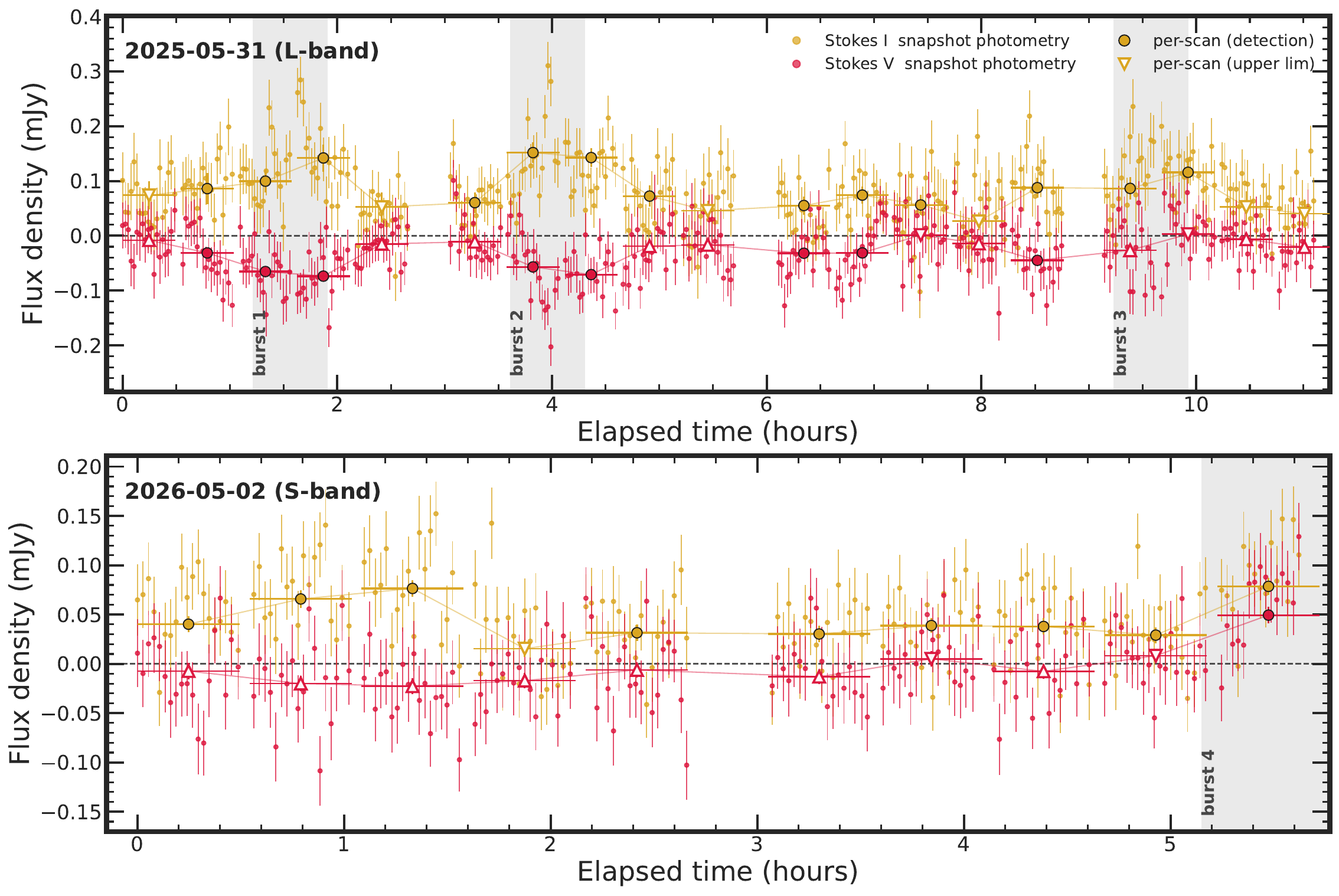}
\caption{\textbf{Rapidly-variable, circularly-polarized radio bursts and inter-burst emission.} Stokes $I$ (yellow) and $V$ (red) light curve of the 2025 May 31 L-band observation (top) and the 2026 May 2 S-band observation (bottom). Small points are snapshot photometry in 96-second bins, while the large outlined markers are per-scan imaging detections and the large triangles are per-scan $3\sigma$ upper limits. Grey vertical bands highlight the recurring, time-variable bursts that dominate the emission. Significant emission is detected outside of the bursts as well, which may represent a quiescent component or overlapping lower amplitude bursts. Positive and negative flux values in circular polarization correspond to right- and left-handed polarization.}
\label{fig:lc}
\end{figure}

\begin{figure}[t]\centering
\includegraphics[width=0.87\linewidth]{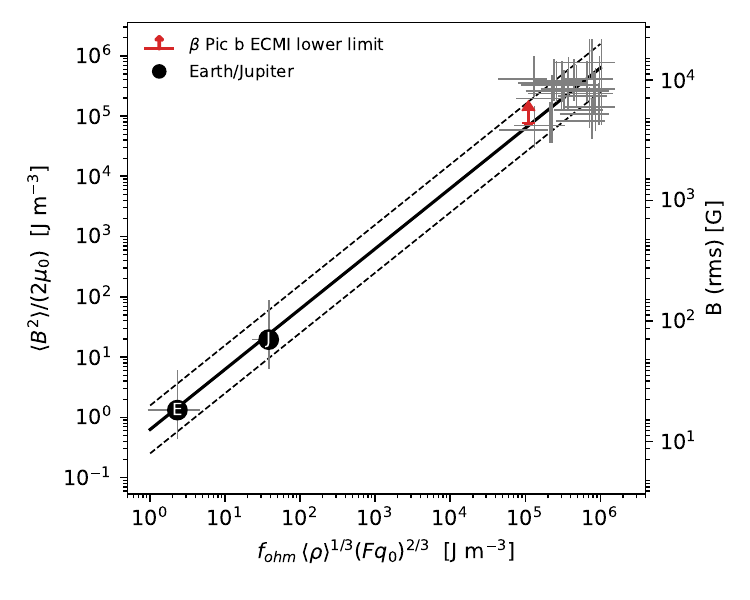}
\caption{\textbf{\bp~b on the magnetic-convective dynamo scaling.} Magnetic energy density (ordinate) versus convected energy density (abscissa) for the \citet{christensen_energy_2009} calibration sample of T~Tauri stars and rapidly rotating M dwarfs (grey error crosses) with the Earth (E) and Jupiter (J) as anchors; the solid line is the scaling relation. 
\bp~b (red) is placed on the abscissa by its measured\cite{reiners_magnetic_2010} luminosity, mass and radius (Extended Data Table~3), while the vertical arrow on the ordinate marks the planetary magnetic field strength lower limit inferred from the highest-observed ECMI frequency in our observations, converted to internal field strength in the same convention as the stellar sample. 
The observed radio emission and properties of \bp~b place it on the correlation.}
\label{fig:christensen}
\end{figure}

The emission from \bp~b is composed of recurring, rapidly-variable bursts with large circular polarization of $\approx 40-70\%$, as well as fainter inter-burst persistent emission (Fig.~\ref{fig:lc}). The strong circular polarization and rapid variability point to electron cyclotron maser instability (ECMI) emission\cite{treumann_electron-cyclotron_2006, dulk_radio_1985} for the bursts; we disfavor plasma emission because it would require implausibly high plasma densities (Methods). The more quiescent component may be either incoherent gyrosynchrotron emission, or potentially composed of overlapping lower-amplitude ECMI bursts. The spectra for all bursts are flat and broadband, with emission present across the full band, as is the case for the continuum emission. The circular polarization handedness is not fixed, with left-handed bursts in the 2025 May 31 L-band epoch and a right-handed burst in the 2026 May 2 S-band epoch (Fig.~\ref{fig:lc}), consistent with auroral ECMI in which the handedness is set by magnetic field orientation and rotation\cite{hallinan_periodic_2007,williams_variable_2017}. For ECMI, the emission frequency is the electron cyclotron frequency, $\nu_{\rm c}={eB}/{2\pi m_{e}c}\approx 2.8\left({B}/{1~{\rm kG}}\right)~{\rm GHz}$\cite{dulk_radio_1985}. The burst detected in the S-band is also the most circularly polarized ($\approx 70\%$) and is detected up to 3.5 GHz, the top of the observing band, therefore requiring a magnetic field of $B\gtrsim 1.25$ kG at the emission site. This constitutes the first direct measurement of magnetic field strength for an exoplanet, and is consistent with dynamo-scaling predictions for a young, massive giant planet\cite{christensen_energy_2009}; see Fig.~\ref{fig:christensen} (Methods). The inferred field strength further rules out stellar emission since the upper limit\cite{zwintz_revisiting_2019} for the host star's dipole field strength is $\lesssim 0.3$ kG (Methods). In light of the radio detection and magnetic field inference, we speculate that weak X-ray emission previously detected from the \bp system and ascribed\cite{gunther_soft_2012} to the star may instead arise from \bp~b, placing it in the regime of the radio/X-ray correlation occupied by UCDs\cite{williams_trends_2014,magaudda_transitions_2024} (Methods).

We attribute the radio emission to magnetosphere-ionosphere coupling at \bp~b. In Jupiter and in UCDs, a fast-spinning magnetosphere loaded with internally-supplied plasma cannot enforce corotation at large distances. The resulting shear drives field-aligned currents that precipitate electrons and drive auroral ECMI, with the radiated power set by the rotation rate, field strength, plasma mass-loading rate and ionospheric conductance\cite{hill_inertial_1979,cowley_origin_2001,nichols_origin_2012}.
\bp~b has the two measurable ingredients for this magnetosphere–ionosphere coupling---rapid rotation ($v\sin i\approx 20$ km s$^{-1}$, $P_{\rm rot}\approx 8-9$~hr from spectroscopy\cite{snellen_fast_2014,landman__2024,janson_deep_2025} and $9.00\pm0.13$~hr from JWST photometry\cite{zhou_photometric_2026}, with a nearly equator-on spin axis) and a kilogauss field as demonstrated here; the origin of the required magnetospheric plasma could include sputtering of the planetary atmosphere from auroral currents\cite{hallinan_magnetospherically_2015} or the capture of debris-disk gas. Rotational modulation is expected from a magnetosphere-ionosphere coupling scenario\cite{nichols_origin_2012}; the $\approx 8$~hr separation between the first and third bursts in L-band (Fig.~\ref{fig:lc}) is comparable to the planetary rotation period, and may be a signature of such modulation.

We note, but disfavor, two alternative drivers for the radio emission. First, stellar wind impinging on the planetary magnetosphere, analogous to how Jupiter's hectometric emission is produced\cite{barrow_solar_1989}, has been predicted to produce radio emission from the \bp planets\cite{katarzynski_search_2016,ashtari_detecting_2022}.
However, scaling the wind-driven radio power to Jupiter's\cite{griesmeier_influence_2005} at \bp~A's mass loss rate\cite{bruhweiler_mass_1991} of $1.1\times10^{-14} M_\odot\,\mathrm{yr}^{-1}$, and at \bp~b's orbital distance and field strength, under-predicts the observed emission power by three orders of magnitude.
The second driver, planet--moon (Io--Jupiter-type) Alfvénic interaction\cite{bigg_influence_1964,goldreich_io_1969} is energetically insufficient. Even maximized over a hypothetical moon's orbital radius and computed for parameters chosen to favor detection, predicted flux from this driver for \bp~b is still more than an order of magnitude below our quiescent emission\cite{katarzynski_search_2016}; existing mass limits also independently exclude the presence of any moons more massive than $\sim$Saturn around \bp~b over most of the dynamically-stable range\cite{macias_first_2026,kenworthy_upper_2026}.
Rotationally-driven coupling accounts for the power and modulation observed in the auroral emission of UCDs\cite{nichols_origin_2012} that \bp~b's bursts closely resemble\cite{kao_strongest_2018}, and many of these UCDs are isolated objects with no companion that can supply an external wind.

\bp~b represents the first case in which radio emission is astrometrically localized to a bona-fide, directly-imaged exoplanet, and angularly separated from its host star. 
Unlike the isolated planetary-mass brown dwarf SIMP J013656.5+093347, which exhibits auroral radio emission\cite{kao_strongest_2018}, \bp~b has a dynamically-measured mass, so that the magnetic field strength implied by the highest-observed ECMI frequency ($\gtrsim1.25$~kG) can be tested against model prediction: it places a massive exoplanet, for the first time, in the regime predicted by a dynamo scaling law calibrated on the solar system planets and low-mass stars\cite{christensen_energy_2009,reiners_magnetic_2010}
(Fig.~\ref{fig:christensen}).
Because the emission should be modulated by rotation, continued radio monitoring can reveal the planet's magnetic obliquity and geometry\cite{kavanagh_unravelling_2024}.
More broadly, seven more directly-imaged giant exoplanets in five other systems within 45~pc lie at separations where the astrometric localization technique used here applies, and a $\sim5\times$ to $7\times$ improvement in instrument sensitivity, expected from next-generation radio observatories, will bring them within reach of detection.

\clearpage

\clearpage
\section*{Methods}
\small

\subsection*{Observations}
We observed the \bp system with MeerKAT on {four} epochs between 2025-02 and 2026-05 (Extended Data Table~1), in the L~band (0.856--1.712~GHz) and S~band (1.75--3.5~GHz), using the standard 4096-channel mode with 8-s integrations. The first L-band epoch was obtained as part of a wider survey in search of quiescent radio emission from UCDs and exoplanets (K.~Ortiz Ceballos et al., in prep.); following the initial detection we obtained a 9.3-hr L-band follow-up through Director's Discretionary Time, and subsequently S-band Open Time. For the S-band campaign we first split the array into two subarrays covering 1.7--2.6 and 2.6--3.5~GHz simultaneously, confirmed emission from the source across the full band, then dedicated an observing block to the 2.6--3.5~GHz sub-band for maximum angular resolution and sensitivity. The bandpass and gain calibrators were J0408$-$6545 and J0538$-$4405 respectively; J0521+1638 (3C138) was added as a polarization calibrator for all but the discovery epoch.

\subsection*{Calibration, imaging and light curves}
The measurement sets for all capture blocks were downloaded from the SARAO archive in both calibrated and uncalibrated formats. The calibrated measurement sets were provided from the SARAO Science Data Processor (SDP) pipeline \texttt{katsdpcal}'s standard mode, and we used WSCLEAN \cite{offringa_wsclean_2014,offringa_optimized_2017} for imaging in Stokes I and V. 
For astrometry, we imaged with Briggs robustness 0.0, because the MeerKAT beam becomes non-Gaussian at higher robustness\cite{ranchod_first_2025,hughes_comprehensive_2025} while lower robustness reduced signal-to-noise and astrometric precision in our tests. A large-field continuum image of the 2026-05-02 S-band observation, produced from this SDP-calibrated measurement set, was the image used for the astrometric registration of the radio source to \bp~b.

Separately, we also calibrated and imaged the uncalibrated measurement sets with the {POLKAT} pipeline\cite{hughes_polkat_2025}, a version of {OXKAT}\cite{heywood_oxkat_2020} modified to implement polarization calibration. {POLKAT} uses CASA\cite{casa_team_casa_2022} for reference calibration, Tricolour\cite{hugo_tricolour_2022} for flagging, QuartiCal\cite{kenyon_africanus_2025} for self-calibration, and WSCLEAN\cite{offringa_wsclean_2014,offringa_optimized_2017} for imaging.
While the SDP products do not undergo a dedicated polarization calibration step, we find that the Stokes~$V$ results agree between the two products. Because MeerKAT has orthogonal linear feeds, Stokes~$V$ occupies the imaginary crosshand correlations\cite{hales_calibration_2017} and can be recovered for a strongly circularly-polarized source even without polarization calibration. Nevertheless, we use the {POLKAT} products for the polarization and time-series analysis due to its use of the polarization calibrator for more reliable Stokes V measurement, and improved flagging in the L-band with Tricolour. Here, we used Briggs robustness +0.25 to improve signal-to-noise ratio (at a mild cost of angular resolution) since we do not use these images for astrometry.

The light curves for Stokes~$I$ and $V$ flux were built by using the {POLKAT} snapshot imaging mode, and then measuring the flux density at the proper- and orbital-motion propagated position of \bp~b for each snapshot. The snapshot images themselves are made by subtracting the steady sky model from the measurement set, dirty imaging at a cadence of 96-second bins, and then restoring the sky model to the snapshot images.
This method measures variability with accurate fluxes but does not clearly distinguish when the source is confidently detected above the noise, so we also use WSCLEAN to image each 30-minute scan individually in the standard continuum fashion in Stokes I and V, and measure the corresponding peak flux or 3$\sigma$ upper limit for each scan for a low-cadence light curve.
We show the light curves for the 2025-05-31 L-band and 2026-05-02 S-band epochs in Fig.~\ref{fig:lc}. 
The remaining epochs (2025-02-15 L-band; 2026-02-20 S-band low \& high) do not show measurable variability due to insufficient SNR per bin, but this lack of variability constrains any rapid bursts in those epochs to within a factor of $\sim2$ (L-band) to $\sim5$ (S-band subarrays) times the measured continuum flux.

\subsection*{Astrometric registration}
We computed the positions of planets~b and c for the 2026-05-02 S-band epoch using \texttt{whereistheplanet}\cite{wang_whereistheplanet_2021}, which propagates the posterior of the orbital solution\cite{lacour_mass_2021} to the observation epoch; for planet~d we adopted its directly-imaged position\cite{sutlieff_direct_2026}. 
We fixed \bp to its proper-motion-corrected {\it Gaia}~DR3 position, and fitted the radio source position with the Aegean source fitter\cite{hancock_compact_2012,hancock_source_2018}. Aegean returns the RA and Dec position uncertainties in addition to the fitted Gaussian position angle. We use these to reconstruct the full $2\times2$ covariance matrix and thus the actual covariance ellipse.
The image absolute registration is anchored on the phase calibrator J0538$-$4405; its VLBI position\cite{charlot_third_2020,petrov_radio_2025} (accurate to $\lesssim0.2$~mas) differs from the tracked position by only 42~mas, and its calibrated visibility phases are flat across the band and stable for each of the 12 phase-referencing scans, confirming a clean phase transfer.
To correct residual frame error (including the calibrator's 42~mas offset) we extracted single-component compact sources in the field with Aegean and tied them to the \textit{Gaia} celestial reference frame.
Our references are {nine} quasars selected from the \textit{Gaia}-CRF3 AGN cross-identification tables, matched within $1''$ to a SNR$\geq15$ radio source with low \textit{Gaia} astrometric excess noise; we also assign a conservative {10}-mas radio--optical offset uncertainty to each reference\cite{gaia_collaboration_gaia_2022}.
One VLBI calibrator in the field from the LBA Calibrator Survey\cite{petrov_second_2019}, LCS~J0542$-$5142 (position accurate to $\sim$5--11~mas)\cite{petrov_radio_2025} was also used, for a total of ten astrometric references (Extended Data Table~2, Extended Data Fig.~1).

We solved for an affine frame-tie correction $\Delta(\mathbf r)=\mathbf t+\mathsf M\mathbf r$, where the translation $\mathbf t$ removes a bulk offset and the $2\times2$ matrix $\mathsf M$ absorbs residual field rotation, differential scale and shear. The fit is a generalized (inverse-covariance-weighted) least-squares solution: each reference enters with its full $2\times2$ tangent-plane covariance $\Sigma_i$, and the design equations are whitened by $\Sigma_i^{-1/2}$ before the solve, so more precise references carry proportionally more weight and the elliptical (rather than scalar) shape of each error is respected. 
$\Sigma_i$ is the sum of the anisotropic radio fitting covariance---proportional to the synthesized-beam major and minor axes and the source SNR\cite{condon_errors_1997}---and the catalog covariance, i.e.\ the {\it Gaia} or VLBI position-error ellipse including its right-ascension--declination correlation, plus the radio--optical offset added in quadrature for the {\it Gaia} sources. 
This follows the affine frame-tie approach previously developed for MeerKAT sub-arcsecond localization\cite{driessen_21_2022,driessen_frb_2024}, but weights each reference by its full anisotropic covariance (beam, catalog and radio--optical terms) rather than using an unweighted transform. 
The fit is statistically satisfactory, with a reduced $\chi^2\approx1.0$ ($\chi^2=14.0$ for 14 degrees of freedom) over the ten references; the residual scatter about the fit is therefore consistent with the adopted per-reference uncertainties.
A translation-only frame-tie is disfavored {($\chi^2/\mathrm{dof}\approx6.3$)}, confirming the need for the rotation, scale and shear terms.

\subsection*{Significance and robustness of the planet assignment}

Given that any offset between the frame-tie corrected radio source position and a candidate counterpart is two-dimensional, we quantify positional coincidence using the Mahalanobis radius $R$, which is the offset in units of the full covariance ellipse.
Radio-source and reference position uncertainties were propagated by Monte Carlo with 40,000 draws. In each draw, we perturb every astrometric reference offset by its full covariance and refit the transform, independently perturb the radio source position by its fitted covariance, and apply the refitted transform to that perturbed position to obtain one realization of the frame-tie corrected position. The comparison positions of the star and the planets are perturbed by their uncertainties, which are the propagated \textit{Gaia} covariance of the star, and also the orbital solution uncertainties for the planets. 
For each comparison hypothesis we compute $R$ from its Monte Carlo offsets, taking their mean as the offset and their scatter as its covariance, so every term above is inside the error budget by construction.
For each of the candidate positions of the host star, planet~b and planet~c, we calculate the probability that measurement noise alone could displace the radio source to the distance of its offset. For two-dimensional Gaussian error, this is $p = P(\chi^2_{2} > R^2) = e^{-R^2/2}$, which we then convert to its equivalent one-dimensional Gaussian significance.
Because noise can scatter in any direction in two dimensions, this quantity is necessarily smaller than $R$ (e.g. $R=4.9$ for the star corresponds to $4.4\sigma$). 
The covariance ellipses of Fig.~\ref{fig:astrometry} show the corrected-position part of this budget; the star and ephemeris terms ($\lesssim$3~mas for the star and planet b, $\lesssim$22~mas for planet c) are accounted for in the reported significances but are too small to draw.
With only the propagated uncertainties of the frame-tie, the corrected radio source position is consistent with planet~b ($R=1.3$, $p=0.44$) and inconsistent with the host star ($R=5.7$, $5.2\sigma$) and
planet~c ($R=6.1$, $5.6\sigma$).
After inflating the covariance ellipse by two empirical systematics (the injection-recovery factors and the ionospheric per-sightline differential;
see below), the assignment is unchanged, with $R=1.1$ ($p=0.53$) for planet~b, the host star excluded at $4.4\sigma$ ($R=4.9$) and planet~c excluded at $4.8\sigma$ ($R=5.3$).
The star and planet~b positions are themselves separated at $R=5.7$ ($5.2\sigma$) of measurement precision even after the systematics covariance inflation ($R=6.5$, $6.1\sigma$ with only the statistical uncertainties), so the two hypotheses are cleanly resolved.
We note that the offset expected from pure measurement scatter in two dimensions follows a Rayleigh distribution with median $R=\sqrt{2\ln2}\approx1.18$, such that the offset between the corrected source position and planet~b is statistically indistinguishable from exact coincidence with the position of the planet.

The identification of planet~b as the source of radio emission is insensitive to the specific choice of astrometric references. 
Under leave-one-out and leave-two-out resampling of the ten references, the corrected position of the radio source remains consistent with planet~b in all 55 fits, with median offsets $\approx${100}~mas from planet~b versus $\approx${434}~mas from the star, and a reference-correction median shift of only $\approx${4}~mas (maximum 29~mas; Extended Data Fig.~4).

\subsection*{Point source injection recovery test}
Because the dominant term in the position uncertainty is the Aegean fitting covariance {of the \bp radio source}, we tested whether it accurately represents the recovery of true source positions given the noise of the 2026-05-02 S-band image.
We injected 160 point sources with the same peak flux of the \bp radio source and the shape of the restoring beam into source-free regions within a $500''$ radius of the image phase center, selecting the regions so that their noise matches that of the vicinity of the \bp radio source to within 15\%; we also enforce a distance of $\geq6$ beams from real detected sources, and reject sites with $\geq4\sigma$ pixels to avoid real background sources.
We recover all injected sources with the identical Aegean method used in the initial fit of the target source.
The recovered positions show a scatter about their injected truth that allows us to empirically measure a fitting covariance (Extended Data Fig.~5).
The empirical scatter of the recovered positions is 1.1$\times$ the Aegean per-axis fitting uncertainty on the \bp~b radio source; this factor is applied in the systematics-inflated covariance shown in Fig.~\ref{fig:astrometry}.

\subsection*{Ionospheric refraction test}
Ionospheric refraction produces a chromatic ($\propto\lambda^2$) position shift, imprinting a dispersive signature in the position scatter of point sources across the field of view. This effect is typically minimal for centimeter bands such as S-band \citep{loi_quantifying_2015}; however, it can be tested directly for a given dataset. Using the {40} brightest compact field sources across 8 sub-bands (2.68--3.45~GHz, every source detected in all sub-bands), we fit source positions with Aegean and measure astrometric scatter with respect to the continuum positions.
Concretely, the position of each source is fit against $\lambda^2$ using inverse-variance-weighted least squares, with each channel weighted by $1/\sigma^2$ where $\sigma$ is the fit positional uncertainty. The fit slope is measured and its uncertainty is then inflated by $\sqrt{\max \left(\chi^2_\nu, 1\right)}$ so any per-source scatter in excess of fitting error is conservatively propagated.
Our radius $\sim1^{\circ}$ field is smaller than a coherent ionospheric patch\cite{cohen_probing_2009}, so a chromatic shift will be nearly common across the field, and we can combine the sources to measure the field-mean dispersive scatter effect, with respect to the wide-band reference frequency $\nu_{\rm ref}={3.06}$\,GHz.
We detect a common-mode shift of $9.8 \pm 2.6$~mas, with a median per-source residual of 32 mas indicating the relevant per-line-of-sight differential (Extended Data Fig.~2). The affine frame-tie absorbs the common-mode shift and any linear gradient, with the higher-order effects responsible for the differential.
We add the differential in quadrature to the radio source position uncertainty as part of the systematics-inflated covariance shown in Fig.~\ref{fig:astrometry}; we implement it as $32/\sqrt{2}\approx23$~mas per axis.
This is conservative: treating an additional per-sightline displacement for the astrometric reference sources as a free parameter of the frame-tie fit, the maximum likelihood is at zero displacement and allows at most 13~mas per axis (95\% confidence). We keep the 23~mas per axis systematic because any ionospheric displacement specific to the target sightline would be invisible to any other test we perform (the frame-tie residuals, the injection test and the jackknife test).

\subsection*{Emission mechanism and magnetic-field estimate}
The astrometric localization assigns the emission to planet~b, so the emitting region cannot exceed the size of the planetary magnetosphere. Adopting a source size of order the planetary radius ($\sim$1.5\,$R_{\rm J}$)\cite{morzinski_magellan_2015} and the brightest burst flux of {307~$\mu$Jy} at L-band results in a brightness temperature lower limit at 1.28 GHz of {$T_b\gtrsim6\times10^{10}$~K}, requiring a nonthermal origin\cite{hallinan_confirmation_2008}.
The high degree of circular polarization ($\sim${40--70}\%), rapid burst timescales and flat spectra identify coherent emission, ruling out incoherent gyrosynchrotron, whose circular polarization and brightness temperature are modest. 
Among coherent mechanisms, we disfavor plasma emission on density grounds: plasma emission radiates at the plasma frequency\cite{dulk_radio_1985} and for the observed frequencies would require electron densities of $n_e\sim9\times10^{9}$---1.5$\times10^{11}$~cm$^{-3}$ (relaxed to only $n_e\sim2\times10^{9}$---4$\times10^{10}$~cm$^{-3}$ for the second harmonic), orders of magnitude above any plausible planetary or brown-dwarf magnetospheric plasma, which instead occupy the low-density ($f_{\rm pe}\ll f_{\rm ce}$) regime in which the electron-cyclotron maser operates\cite{treumann_electron-cyclotron_2006,hallinan_confirmation_2008}.
ECMI requires only a kilogauss field, consistent with expectations for a young, massive giant planet from dynamo scaling\cite{christensen_energy_2009}; we therefore conclude ECMI is the emission mechanism. Under the fundamental-ECMI interpretation, $\nu_{\rm c}={eB}/{2\pi m_{e}c}\approx 2.8\left({B}/{1~{\rm kG}}\right)~{\rm GHz}$\cite{dulk_radio_1985} maps the highest emission frequency to the field strength at the source: 1.7~GHz $\rightarrow$ $\gtrsim$600~G for the L-band epochs, and 3.5~GHz $\rightarrow$ $\gtrsim$1.25~kG for the S-band epochs. 
Independent dynamo scaling supports a kilogauss field for \bp~b. Applying the energy-flux scaling law\cite{christensen_energy_2009,reiners_magnetic_2010} to the planet's measured luminosity, mass and radius (Extended Data Table~3) predicts a 
mean dynamo surface field of $\approx$1.2~kG, corresponding to a polar dipole of $\approx$0.8~kG.
Our highest observational limit exceeds that dipole strength by a factor of 1.6, expected for an ECMI constraint: the cyclotron frequency samples the local field strength at the point of emission; for the case of Jupiter, the field at the Io flux tube's auroral footprint\cite{connerney_new_2022} exceeds the dipole strength by a factor of $\approx$2. 
{Note that Fig.~\ref{fig:christensen} is expressed as the root-mean-square dynamo-region field\cite{christensen_energy_2009} $\langle B\rangle \approx 3.5\,B_{\rm s}$, so the predicted 1.2~kG surface field becomes $\langle B\rangle\approx4.2$~kG and our 1.25~kG ECMI limit becomes $\langle B\rangle\approx4.4$~kG in the same convention.}

\subsection*{Excluded stellar-origin alternatives} 

We exclude the debris disk as a plausible source of the radio emission. Thermal emission from the disk is negligible in the centimeter bands; the (sub)millimeter spectral index\cite{ricci_atca_2015} extrapolated to our observing frequencies predicts emission at nanojansky levels.
The host star is also excluded on two independent grounds. First and decisively, the astrometric mismatch with the location of the radio source excludes the star with {4.4$\sigma$} significance. Second, the star is magnetically quiet. HARPSpol spectropolarimetry failed to detect a 
large-scale surface magnetic field ($\langle B_z\rangle=-14\pm20$~G),
and from oblique-dipole modeling excludes a large-scale dipole with polar strength $B_{\rm pol}\gtrsim 300$~G at 90\% detection probability (down to 120~G at 50\%)\cite{zwintz_revisiting_2019}.
This is a quarter of the $\gtrsim1.25$~kG field strength required to produce ECMI at the observed frequencies, and an order of magnitude below the typical multi-kilogauss fields of the radio-bright magnetic Ap/Bp stars, which are also hotter\cite{leto_scaling_2021} (spectral types A2-B2) than the star (spectral type A6V). Consistent with this nondetection, $\beta$~Pictoris is not chemically peculiar\cite{saffe_chemical_2021}, and thus not of the Ap/Bp class of stars that host stable fossil fields. 

No physical mechanism known to cause radio emission in early-type stars can explain the observed emission. 
Chromospheric free-free emission is ruled out from the brightness temperature constraint; reproducing even the quiescent flux with a source the size of the stellar disk would require at least $T_{\rm b}\sim10^{7}$~K from the chromosphere, three orders of magnitude above the chromospheric temperature of comparable A-type stars\cite{white_first_2021}.
Incoherent gyrosynchrotron from a centrifugal stellar magnetosphere\cite{owocki_centrifugal_2022} does not reach the measured $40\%$--$70\%$ circular polarization fraction\cite{leto_scaling_2021}.
And given the $\lesssim 300$~G constraint on the stellar dipole, coherent ECMI as seen in main-sequence radio pulse emitters\cite{das_discovery_2022} is impossible at the S-band frequencies. The dipole is strongest at the surface, so no region of the stellar magnetosphere reaches the $\gtrsim1.25$~kG required at the fundamental, or even the $\gtrsim0.63$~kG required at the second harmonic.
In the pulse emitters themselves the emission originates
$\gtrsim0.6\,R_*$ above the surface\cite{das_discovery_2022}, where any stellar dipole field would be even weaker.

The detection of soft X-rays from the \bp system is consistent with either a stellar or planetary origin. 
Placed on the empirical radio/X-ray (G\"udel--Benz) relation\cite{guedel_x-raymicrowave_1993,benz_x-raymicrowave_1994} using a soft X-ray luminosity of $\log L_X = 26.5$ (0.2–2.0 keV), from the 1.1 MK fit to joint XMM and Chandra data\cite{hempel_detection_2005,gunther_soft_2012}, \bp~b lies far into the radio-overluminous regime occupied by UCDs (Extended Data Fig.~3).
G\"udel--Benz placement assumes a shared origin for the radio and X-ray emissions; the offset from the relation is also naturally explained if the radio emission is planetary and the X-ray emission is stellar. The \textit{Chandra} PSF and absolute astrometry cannot resolve the host star from its planets, and its $T_{\rm eff} = 8090$~K sits essentially at the effective temperature boundary above which single A-type stars cease to exhibit X-ray emission\cite{schroder_x-ray_2007,gunther_coronal_2022}. We therefore speculate that the X-rays may, like the radio emission, arise from planet~b.

\subsection*{Data availability}
\vspace{-2mm}
The MeerKAT visibility data are stored in the SARAO archive (\url{https://archive.sarao.ac.za}) under capture block IDs 1739610086, 1748669413, 1771595870, 1771595871, and 1777713316. Some capture blocks may still be under proprietary period pending journal publication.
\vspace{-2mm}
\subsection*{Code availability}
\vspace{-2mm}
The end-to-end astrometric registration code will be made available upon publication.

\normalsize

\clearpage
\section*{Extended Data}

\begin{table}[htbp]\centering\small
\noindent\textbf{Extended Data Table 1 $\mid$ MeerKAT observation summary.}\par\medskip
\resizebox{\linewidth}{!}{\begin{tabular}{llcccccc}
\hline
Obs. & Capture block & Time range (UTC) & Freq.\ (GHz) & BW (GHz) & Pol.\ cal. & $N_{\rm ant}$ & $t_{\rm int}$ (hr)\\
\hline
2025-02-15 L-band & 1739610086 & 12:34--16:18 & 0.856--1.712 & 0.856 & -- & 63 & 0.744\\
2025-05-31 L-band & 1748669413 & 05:31--17:19 & 0.856--1.712 & 0.856 & J0521+1638 & 59 & 9.308\\
2026-02-20 S-band (low, subarray) & 1771595870 & 14:00--20:22 & 1.750--2.625 & 0.875 & J0521+1638 & 28 & 4.937\\
2026-02-20 S-band (high, subarray) & 1771595871 & 14:00--20:23 & 2.625--3.500 & 0.875 & J0521+1638 & 29 & 4.942\\
2026-05-02 S-band (high) & 1777713316 & 09:16--15:40 & 2.625--3.500 & 0.875 & J0521+1638 & 60 & 4.875\\
\hline
\end{tabular}}
\end{table}

\begin{table}[htbp]\centering\small
\noindent\textbf{Extended Data Table 2 $\mid$ Astrometric reference sources.}\par\medskip
\setlength{\tabcolsep}{4pt}
\resizebox{\linewidth}{!}{\begin{tabular}{lcccccccc}
\hline
Identifier & Type & $\rho$ (deg) & PA (deg) & $S_{\rm pk}$ ($\mu$Jy) & SNR & $\theta_{\rm maj}/\theta_{\rm bm}$ & $|\Delta|$ (mas) & $\sigma_{\rm fit},\sigma_{\rm cat}$ (mas)\\
\hline
LCS J0542$-$5142 & VLBI & 1.003 & 229 & 37 & 14.5 & 1.26 & 175 & 110,\,6.9\\
\gaia\,4792775209861961344 & CRF3/AGN & 0.062 & 24 & 56 & 19.7 & 1.05 & 213 & 77,\,0.2\\
\gaia\,4792770090259838720 & CRF3/AGN & 0.173 & 224 & 42 & 16.6 & 1.01 & 85 & 86,\,0.3\\
\gaia\,4792785341688756992 & CRF3/AGN & 0.195 & 87 & 101 & 38.7 & 1.00 & 17 & 37,\,0.3\\
\gaia\,4792764450967618048 & CRF3/AGN & 0.226 & 183 & 62 & 24.1 & 1.02 & 118 & 60,\,0.2\\
\gaia\,4792749818015547392 & CRF3/AGN & 0.458 & 154 & 423 & 161.2 & 1.00 & 48 & 10,\,0.4\\
\gaia\,4793535277340873344 & CRF3/AGN & 0.467 & 312 & 121 & 48.3 & 1.00 & 130 & 30,\,0.5\\
\gaia\,4792909174185830912 & CRF3/AGN & 0.602 & 353 & 49 & 19.5 & 1.09 & 247 & 76,\,0.3\\
\gaia\,4792898282149185280 & CRF3/AGN & 0.692 & 26 & 307 & 116.0 & 1.23 & 131 & 14,\,0.3\\
\gaia\,4793178691973522688 & CRF3/AGN & 0.760 & 254 & 181 & 71.2 & 1.12 & 131 & 21,\,0.1\\
\hline
\end{tabular}}\par\medskip
\begin{minipage}{0.95\linewidth}\footnotesize
$\rho$, PA: separation and position angle (east-of-north) from the phase center. $S_{\rm pk}$: Aegean peak. $\theta_{\rm maj}/\theta_{\rm bm}$: fitted major axis / beam (compactness). $|\Delta|$: radio$-$catalog offset (radio$-${\it Gaia} for quasars; radio$-$VLBI for the VLBI source). $\sigma_{\rm fit}$: measured Aegean position uncertainty; $\sigma_{\rm cat}$: catalog uncertainty.
\end{minipage}
\end{table}

\begin{table}[htbp]\centering\small
\noindent\textbf{Extended Data Table 3 $\mid$ System parameters for \bp and its planets.}\par\medskip
\begin{tabular}{lllc}
\hline
Object & Parameter & Value & Ref.\\
\hline
$\beta$~Pic~A & Spectral type & A6V & \citenum{gray_contributions_2006}\\
 & Distance (pc) & $19.63\pm0.06$ & \citenum{gaia_collaboration_gaia_2023}\\
 & Age (Myr) & $23\pm8$ & \citenum{lee_revisiting_2024}\\
 & $T_{\rm eff}$ (K) & $8090\pm59$ & \citenum{zwintz_revisiting_2019}\\
 & Mass ($M_\odot$) & $1.789^{+0.024}_{-0.027}$ & \citenum{sutlieff_direct_2026}\\
 & Mass-loss rate ($M_\odot\,\mathrm{yr}^{-1}$) & $1.1\times10^{-14}$ & \citenum{bruhweiler_mass_1991}\\
 & Radius ($R_\odot$) & $1.497\pm0.025$ & \citenum{zwintz_revisiting_2019}\\
\hline
\bp~b & Semi-major axis (AU) & $9.93\pm0.03$ & \citenum{lacour_mass_2021}\\
 & Mass ($M_{\rm J}$) & $11.90^{+2.93}_{-3.04}$ & \citenum{lacour_mass_2021}\\
& Radius ($R_{\rm J}$) & $1.45\pm0.02$ & \citenum{morzinski_magellan_2015}\\
& $\log(L_{\rm bol}/L_\odot)$ & $-3.78\pm0.03$ & \citenum{morzinski_magellan_2015}\\
 & $T_{\rm eff}$ (K) & $1742\pm10$ & \citenum{gravity_collaboration_peering_2020}\\
 & $v\sin i$ (km\,s$^{-1}$) & $20.36\pm0.31$ & \citenum{janson_deep_2025}\\
\hline
$\beta$~Pic c & Semi-major axis (AU) & $2.68\pm0.02$ & \citenum{lacour_mass_2021}\\
 & Mass ($M_{\rm J}$) & $8.89\pm0.75$ & \citenum{lacour_mass_2021}\\
 & $T_{\rm eff}$ (K) & $1250\pm50$ & \citenum{nowak_direct_2020}\\
\hline
$\beta$~Pic d & Semi-major axis (AU) & $26.0^{+2.2}_{-6.1}$ & \citenum{sutlieff_direct_2026}\\
 & Mass ($M_{\rm J}$) & $2.4\pm0.6$ & \citenum{sutlieff_direct_2026}\\
 & $T_{\rm eff}$ (K) & $600^{+45}_{-60}$ & \citenum{sutlieff_direct_2026}\\
\hline
\end{tabular}
\end{table}

\begin{figure}[htbp]\centering
\vspace{2cm}
\includegraphics[width=0.9\linewidth]{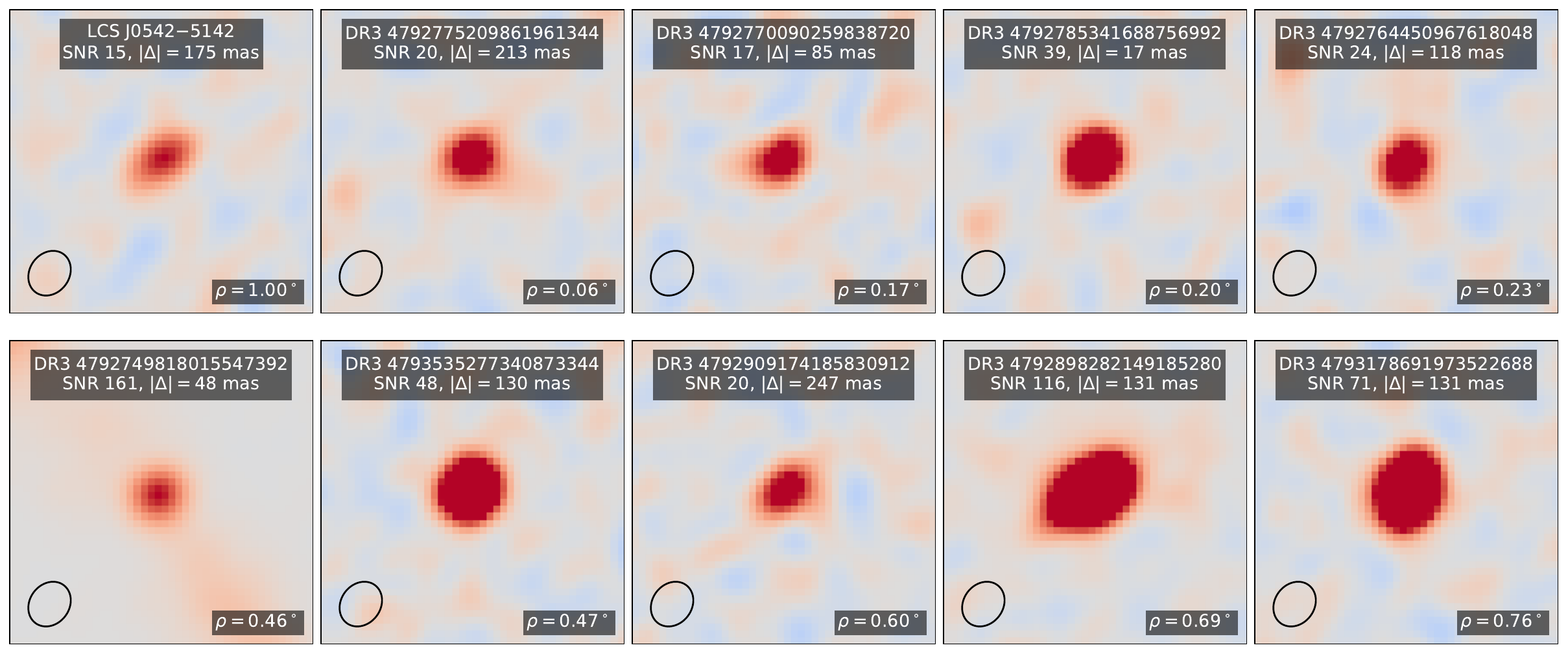}\\[4pt]
\begin{minipage}{0.95\linewidth}\footnotesize
\vspace{0.5cm}
\textbf{Extended Data Fig. 1 $\mid$ Astrometric reference cutouts.} 2026-05-02 S-band image cutouts of the frame-tie reference sources, each labeled with its identifier, SNR, radio$-$catalog offset ($|\Delta|$) and separation from the phase center ($\rho$).
\vspace{2cm}
\end{minipage}
\end{figure}

\begin{figure}[htbp]\centering
\includegraphics[width=0.7\linewidth]{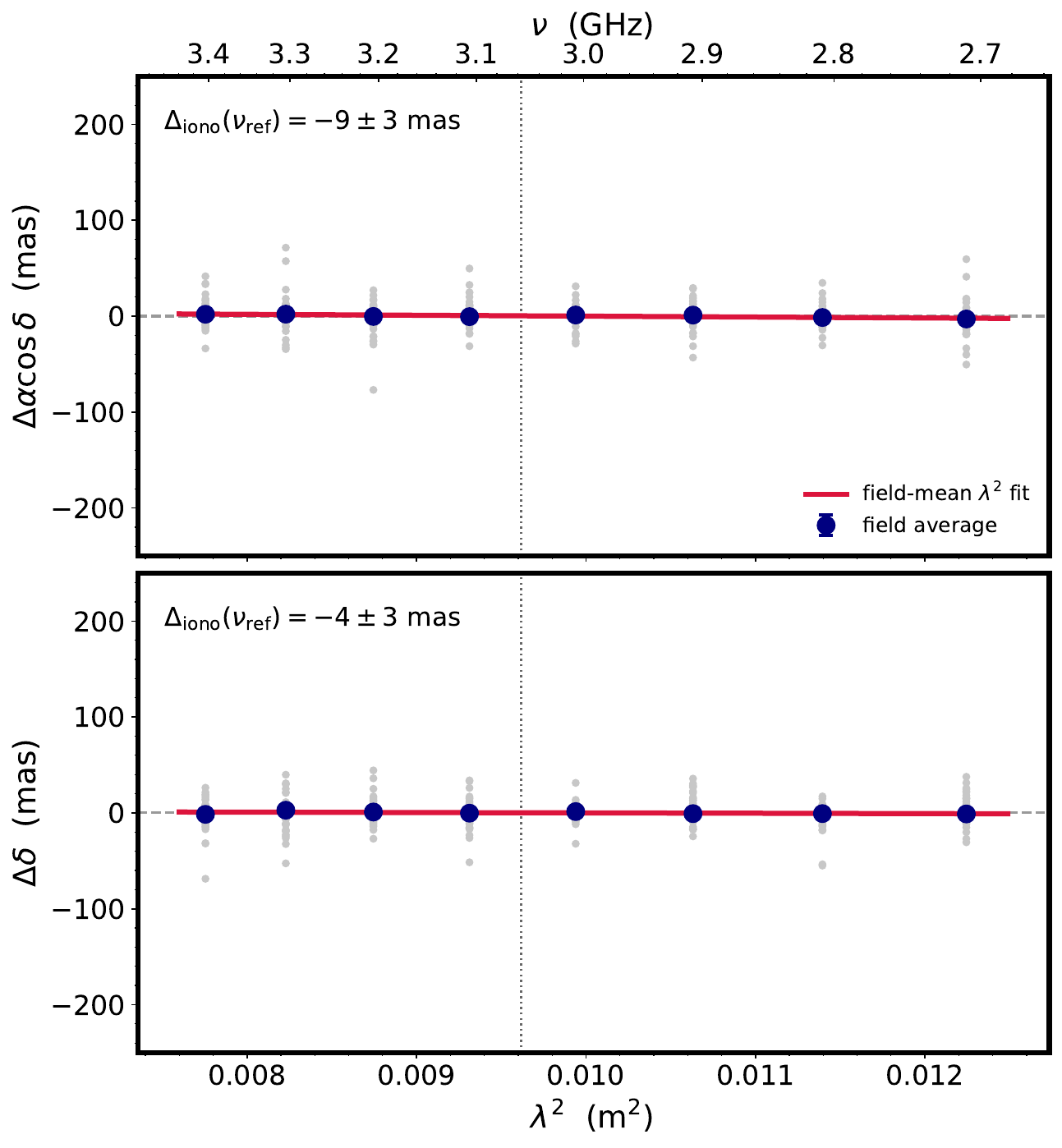}\\[4pt]
\begin{minipage}{0.95\linewidth}\footnotesize
\textbf{Extended Data Fig. 2 $\mid$ Ionospheric refraction $\lambda^2$ test.} Field-averaged fitted position scatter of the brightest compact field sources versus $\lambda^2$ (blue), with individual per-source offsets (grey); the inverse-variance-weighted mean trend (red) shows a common-mode chromatic shift of $9.8\pm2.6$ mas, which is a bulk offset common across the field and absorbed by the affine frame tie. The scatter of the individual sources (median 32~mas) is the per-line-of-sight differential not absorbed by the frame-tie. 
\end{minipage}
\end{figure}

\begin{figure}[htbp]\centering
\includegraphics[width=0.8\linewidth]{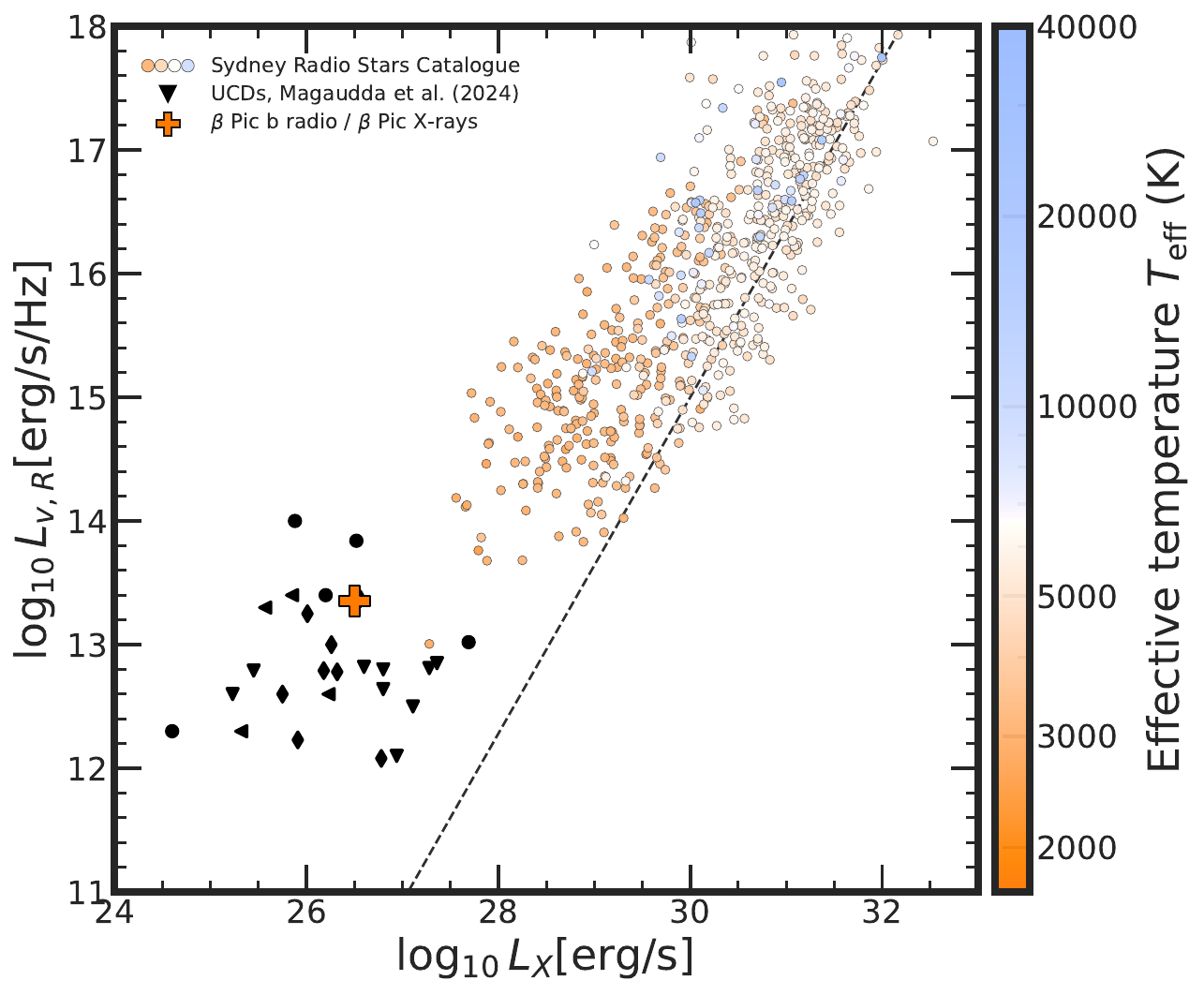}\\[4pt]
\begin{minipage}{0.95\linewidth}\footnotesize
\textbf{Extended Data Fig. 3 $\mid$ \bp~b on the G\"udel-Benz relation.}
The Sydney Radio Stars Catalogue sample of radio stars, colored by effective temperature, matched to counterpart ROSAT detections\cite{driessen_sydney_2024,freund_stellar_2022} (0.1--2.4~keV
band, circles), and a sample of UCDs with eROSITA (0.2--2.0~keV band) and radio measurements\cite{magaudda_transitions_2024} (black; circles, detected in both bands; downward arrows, radio upper limits; leftward arrows, X-ray upper limits; diamonds, upper limits in both bands). The dashed line shows the canonical radio/X-ray relation\cite{guedel_x-raymicrowave_1993,benz_x-raymicrowave_1994,williams_trends_2014}.
The cross marks the position of \bp~b from the system's soft X-ray luminosity $\log L_{\rm X}=26.5$ (0.2--2.0~keV)\cite{gunther_soft_2012} and the quiescent S-band radio flux of 48$\mu$Jy, colored with the planet's effective temperature.
\bp~b lies well into the radio-overluminous UCD regime, 3.1 dex offset from the canonical relation.
\end{minipage}
\end{figure}

\begin{figure}[htbp]\centering
\includegraphics[width=0.72\linewidth]{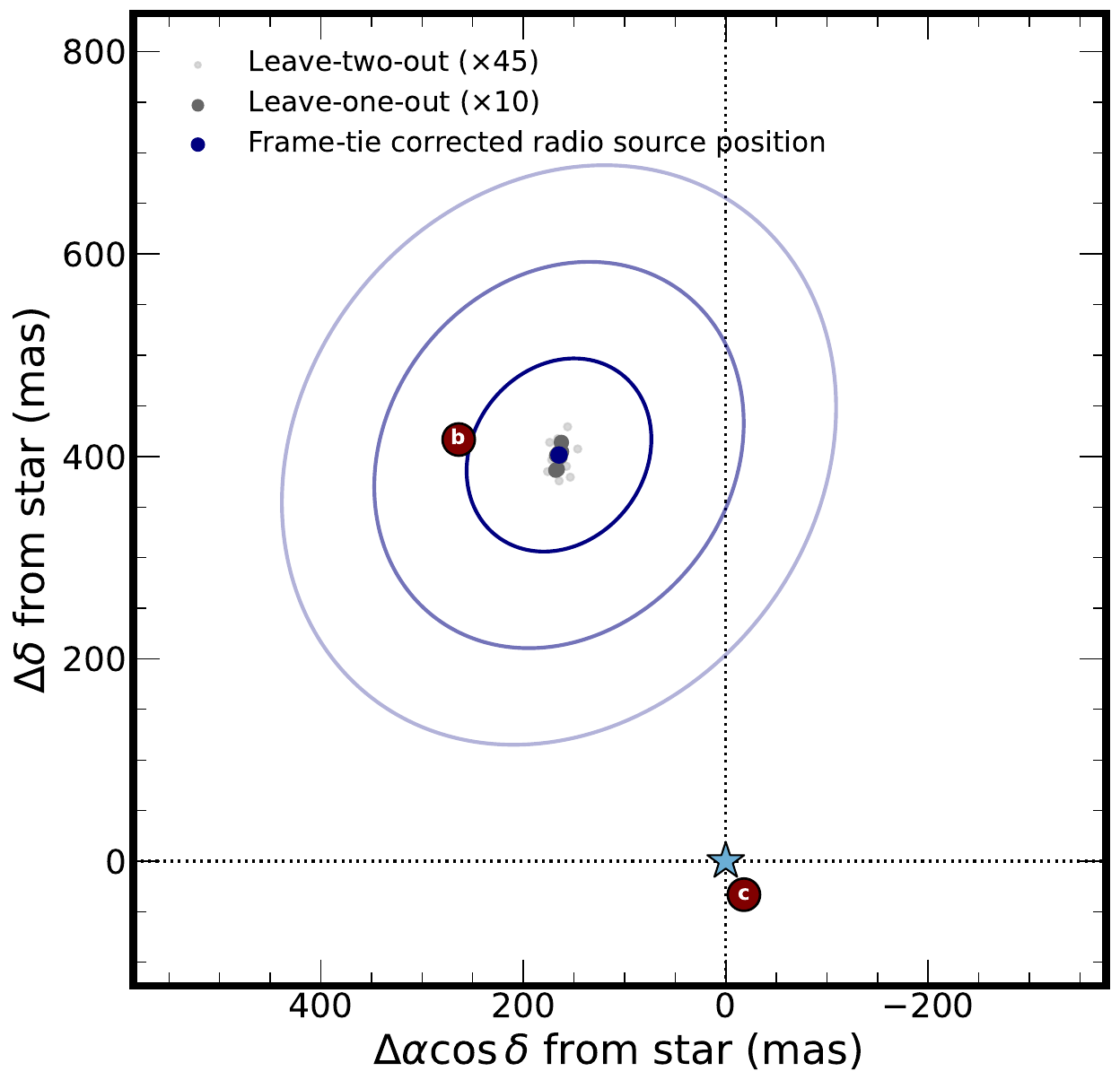}\\[4pt]
\begin{minipage}{0.95\linewidth}\footnotesize
\textbf{Extended Data Fig. 4 $\mid$ Jackknife stability of the localization.} The frame-tie--corrected radio position (blue, with 1,2,3$\sigma$ covariance ellipses) and every leave-one-out and leave-two-out affine frame-tie solution using a reduced number of astrometric references, shown in the same frame as Fig.~\ref{fig:astrometry}. All jackknife solutions cluster at the same radio source position and remain $R=4.6$--$5.0$ from the star, confirming the assignment is not driven by any single reference.
\end{minipage}
\end{figure}

\begin{figure}[htbp]\centering
\includegraphics[width=0.72\linewidth]{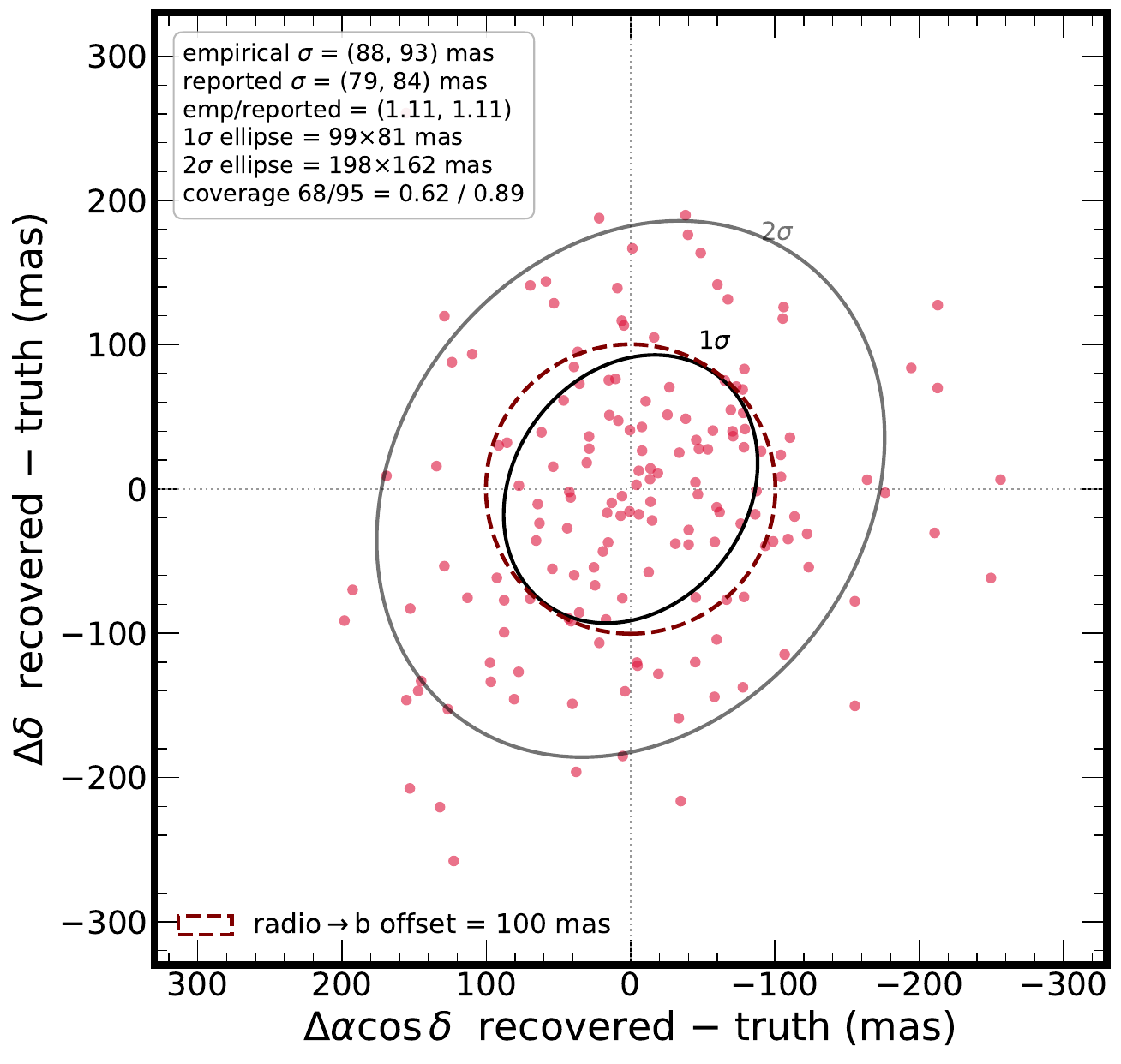}\\[4pt]
\begin{minipage}{0.95\linewidth}\footnotesize
\textbf{Extended Data Fig. 5 $\mid$ Point source injection recovery test.} 160 artificial point sources with the same flux density as \bp~b, injected into the image noise and retrieved with Aegean. Solid lines show the 1$\sigma$ and 2$\sigma$ scatter of the distribution, and the dashed circle shows the offset between planet b and the radio-source position in the astrometric registration.
\end{minipage}
\end{figure}

\clearpage
\bibliographystyle{unsrtnat}
\bibliography{S_submit}

\begin{thebibliography}{97}
\providecommand{\natexlab}[1]{#1}
\providecommand{\url}[1]{\texttt{#1}}
\expandafter\ifx\csname urlstyle\endcsname\relax
  \providecommand{\doi}[1]{doi: #1}\else
  \providecommand{\doi}{doi: \begingroup \urlstyle{rm}\Url}\fi

\bibitem[Brain et~al.(2024)Brain, Kao, and O'Rourke]{brain_exoplanet_2024}
David~A. Brain, Melodie~M. Kao, and Joseph~G. O'Rourke.
\newblock Exoplanet {Magnetic} {Fields}.
\newblock \emph{Reviews in Mineralogy and Geochemistry}, 90:\penalty0 375--410, July 2024.
\newblock \doi{10.2138/rmg.2024.90.11}.

\bibitem[Treumann(2006)]{treumann_electron-cyclotron_2006}
Rudolf~A. Treumann.
\newblock The electron-cyclotron maser for astrophysical application.
\newblock \emph{Astronomy and Astrophysics Review}, 13:\penalty0 229--315, August 2006.
\newblock ISSN 0935-4956.
\newblock \doi{10.1007/s00159-006-0001-y}.

\bibitem[Burke and Franklin(1955)]{burke_observations_1955}
B.~F. Burke and K.~L. Franklin.
\newblock Observations of a {Variable} {Radio} {Source} {Associated} with the {Planet} {Jupiter}.
\newblock \emph{Journal of Geophysical Research}, 60:\penalty0 213--217, June 1955.
\newblock ISSN 0148-0227.
\newblock \doi{10.1029/JZ060i002p00213}.

\bibitem[Berger et~al.(2001)Berger, Ball, Becker, Clarke, Frail, et~al.]{berger_discovery_2001}
E.~Berger, S.~Ball, K.~M. Becker, M.~Clarke, D.~A. Frail, et~al.
\newblock Discovery of radio emission from the brown dwarf {LP944}-20.
\newblock \emph{Nature}, 410:\penalty0 338--340, March 2001.
\newblock ISSN 0028-0836.
\newblock \doi{10.1038/35066514}.

\bibitem[Kao et~al.(2018)Kao, Hallinan, Pineda, Stevenson, and Burgasser]{kao_strongest_2018}
Melodie~M. Kao, Gregg Hallinan, J.~Sebastian Pineda, David Stevenson, and Adam Burgasser.
\newblock The {Strongest} {Magnetic} {Fields} on the {Coolest} {Brown} {Dwarfs}.
\newblock \emph{The Astrophysical Journal Supplement Series}, 237:\penalty0 25, August 2018.
\newblock ISSN 0067-0049.
\newblock \doi{10.3847/1538-4365/aac2d5}.

\bibitem[Vedantham et~al.(2020)Vedantham, Callingham, Shimwell, Tasse, Pope, et~al.]{vedantham_coherent_2020}
H.~K. Vedantham, J.~R. Callingham, T.~W. Shimwell, C.~Tasse, B.~J.~S. Pope, et~al.
\newblock Coherent radio emission from a quiescent red dwarf indicative of star-planet interaction.
\newblock \emph{Nature Astronomy}, 4:\penalty0 577--583, February 2020.
\newblock ISSN 2397-3366.
\newblock \doi{10.1038/s41550-020-1011-9}.

\bibitem[Callingham et~al.(2024)Callingham, Pope, Kavanagh, Bellotti, Daley-Yates, et~al.]{callingham_radio_2024}
J.~R. Callingham, B.~J.~S. Pope, R.~D. Kavanagh, S.~Bellotti, S.~Daley-Yates, et~al.
\newblock Radio signatures of star-planet interactions, exoplanets and space weather.
\newblock \emph{Nature Astronomy}, 8:\penalty0 1359--1372, November 2024.
\newblock ISSN 2397-3366.
\newblock \doi{10.1038/s41550-024-02405-6}.

\bibitem[Hess and Zarka(2011)]{hess_modeling_2011}
S.~L.~G. Hess and P.~Zarka.
\newblock Modeling the radio signature of the orbital parameters, rotation, and magnetic field of exoplanets.
\newblock \emph{Astronomy and Astrophysics}, 531:\penalty0 A29, July 2011.
\newblock ISSN 0004-6361.
\newblock \doi{10.1051/0004-6361/201116510}.

\bibitem[Dulk et~al.(1992)Dulk, Lecacheux, and Leblanc]{dulk_complete_1992}
G.~A. Dulk, A.~Lecacheux, and Y.~Leblanc.
\newblock The complete polarization state of a storm of millisecond bursts from {Jupiter}.
\newblock \emph{Astronomy and Astrophysics}, 253:\penalty0 292--306, January 1992.
\newblock ISSN 0004-6361.

\bibitem[Zarka(1998)]{zarka_auroral_1998}
Philippe Zarka.
\newblock Auroral radio emissions at the outer planets: {Observations} and theories.
\newblock \emph{Journal of Geophysical Research}, 103:\penalty0 20159--20194, September 1998.
\newblock ISSN 0148-0227.
\newblock \doi{10.1029/98JE01323}.

\bibitem[Hallinan et~al.(2007)Hallinan, Bourke, Lane, Antonova, Zavala, et~al.]{hallinan_periodic_2007}
G.~Hallinan, S.~Bourke, C.~Lane, A.~Antonova, R.~T. Zavala, et~al.
\newblock Periodic {Bursts} of {Coherent} {Radio} {Emission} from an {Ultracool} {Dwarf}.
\newblock \emph{The Astrophysical Journal Letters}, 663:\penalty0 L25--L28, July 2007.
\newblock ISSN 0004-637X.
\newblock \doi{10.1086/519790}.

\bibitem[Berger et~al.(2009)Berger, Rutledge, Phan-Bao, Basri, Giampapa, et~al.]{berger_periodic_2009}
E.~Berger, R.~E. Rutledge, N.~Phan-Bao, G.~Basri, M.~S. Giampapa, et~al.
\newblock Periodic {Radio} and {Hα} {Emission} from the {L} {Dwarf} {Binary} {2MASSW} {J0746425}+200032: {Exploring} the {Magnetic} {Field} {Topology} and {Radius} {Of} {An} {L} {Dwarf}.
\newblock \emph{The Astrophysical Journal}, 695:\penalty0 310--316, April 2009.
\newblock ISSN 0004-637X.
\newblock \doi{10.1088/0004-637X/695/1/310}.

\bibitem[Hallinan et~al.(2015)Hallinan, Littlefair, Cotter, Bourke, Harding, et~al.]{hallinan_magnetospherically_2015}
G.~Hallinan, S.~P. Littlefair, G.~Cotter, S.~Bourke, L.~K. Harding, et~al.
\newblock Magnetospherically driven optical and radio aurorae at the end of the stellar main sequence.
\newblock \emph{Nature}, 523\penalty0 (7562):\penalty0 568--571, July 2015.
\newblock ISSN 0028-0836.
\newblock \doi{10.1038/nature14619}.

\bibitem[Williams et~al.(2017)Williams, Gizis, and Berger]{williams_variable_2017}
P.~K.~G. Williams, J.~E. Gizis, and E.~Berger.
\newblock Variable and {Polarized} {Radio} {Emission} from the {T6} {Brown} {Dwarf} {WISEP} {J112254}.73+255021.5.
\newblock \emph{The Astrophysical Journal}, 834:\penalty0 117, January 2017.
\newblock ISSN 0004-637X.
\newblock \doi{10.3847/1538-4357/834/2/117}.

\bibitem[Villadsen and Hallinan(2019)]{villadsen_ultra-wideband_2019}
Jackie Villadsen and Gregg Hallinan.
\newblock Ultra-wideband {Detection} of 22 {Coherent} {Radio} {Bursts} on {M} {Dwarfs}.
\newblock \emph{The Astrophysical Journal}, 871:\penalty0 214, February 2019.
\newblock ISSN 0004-637X.
\newblock \doi{10.3847/1538-4357/aaf88e}.

\bibitem[Callingham et~al.(2021)Callingham, Vedantham, Shimwell, Pope, Davis, et~al.]{callingham_population_2021}
J.~R. Callingham, H.~K. Vedantham, T.~W. Shimwell, B.~J.~S. Pope, I.~E. Davis, et~al.
\newblock The population of {M} dwarfs observed at low radio frequencies.
\newblock \emph{Nature Astronomy}, 5:\penalty0 1233--1239, December 2021.
\newblock ISSN 2397-3366.
\newblock \doi{10.1038/s41550-021-01483-0}.

\bibitem[Cendes et~al.(2022)Cendes, Williams, and Berger]{cendes_pilot_2022}
Y.~Cendes, P.~K.~G. Williams, and E.~Berger.
\newblock A {Pilot} {Radio} {Search} for {Magnetic} {Activity} in {Directly} {Imaged} {Exoplanets}.
\newblock \emph{The Astronomical Journal}, 163:\penalty0 15, January 2022.
\newblock ISSN 0004-6256.
\newblock \doi{10.3847/1538-3881/ac32c8}.

\bibitem[Shiohira et~al.(2024)Shiohira, Fujii, Kita, Kimura, Terada, et~al.]{shiohira_search_2024}
Yuta Shiohira, Yuka Fujii, Hajime Kita, Tomoki Kimura, Yuka Terada, et~al.
\newblock A search for auroral radio emission from β {Pictoris} b.
\newblock \emph{Monthly Notices of the Royal Astronomical Society}, 528:\penalty0 2136--2144, February 2024.
\newblock ISSN 0035-8711.
\newblock \doi{10.1093/mnras/stad3990}.

\bibitem[Turner et~al.(2021)Turner, Zarka, Grießmeier, Lazio, Cecconi, et~al.]{turner_search_2021}
Jake~D. Turner, Philippe Zarka, Jean-Mathias Grießmeier, Joseph Lazio, Baptiste Cecconi, et~al.
\newblock The search for radio emission from the exoplanetary systems 55 {Cancri}, υ {Andromedae}, and τ {Boötis} using {LOFAR} beam-formed observations.
\newblock \emph{Astronomy and Astrophysics}, 645:\penalty0 A59, January 2021.
\newblock ISSN 0004-6361.
\newblock \doi{10.1051/0004-6361/201937201}.

\bibitem[Zhang et~al.(2025)Zhang, Zarka, Girard, Tasse, Loh, et~al.]{zhang_circularly_2025}
X.~Zhang, P.~Zarka, J.~N. Girard, C.~Tasse, A.~Loh, et~al.
\newblock A circularly polarized low-frequency radio burst from the exoplanetary system {HD} 189733.
\newblock \emph{Astronomy and Astrophysics}, 700:\penalty0 A140, August 2025.
\newblock ISSN 0004-6361.
\newblock \doi{10.1051/0004-6361/202555209}.

\bibitem[Turner et~al.(2026)Turner, Zarka, Grießmeier, Louis, Zhang, et~al.]{turner_tentative_2026}
Jake~D. Turner, Philippe Zarka, Jean-Mathias Grießmeier, Corentin~K. Louis, Xiang Zhang, et~al.
\newblock Tentative detection of circularly polarized bursty radio emissions from the {HD} 189733 exoplanetary system using {NenuFAR} beamformed observations.
\newblock arXiv:2607.08910, July 2026.

\bibitem[{Gaia Collaboration} et~al.(2023){Gaia Collaboration}, Vallenari, Brown, Prusti, de~Bruijne, et~al.]{gaia_collaboration_gaia_2023}
{Gaia Collaboration}, A.~Vallenari, A.~G.~A. Brown, T.~Prusti, J.~H.~J. de~Bruijne, et~al.
\newblock Gaia {Data} {Release} 3. {Summary} of the content and survey properties.
\newblock \emph{Astronomy and Astrophysics}, 674:\penalty0 A1, June 2023.
\newblock ISSN 0004-6361.
\newblock \doi{10.1051/0004-6361/202243940}.

\bibitem[Lee et~al.(2024)Lee, Gaidos, van Saders, Feiden, and Gagné]{lee_revisiting_2024}
Rena~A. Lee, Eric Gaidos, Jennifer van Saders, Gregory~A. Feiden, and Jonathan Gagné.
\newblock Revisiting the membership, multiplicity, and age of the {Beta} {Pictoris} {Moving} {Group} in the {Gaia} era.
\newblock \emph{Monthly Notices of the Royal Astronomical Society}, 528:\penalty0 4760--4774, March 2024.
\newblock ISSN 0035-8711.
\newblock \doi{10.1093/mnras/stae007}.

\bibitem[Smith and Terrile(1984)]{smith_circumstellar_1984}
B.~A. Smith and R.~J. Terrile.
\newblock A {Circumstellar} {Disk} around β {Pictoris}.
\newblock \emph{Science}, 226:\penalty0 1421--1424, December 1984.
\newblock ISSN 0036-8075.
\newblock \doi{10.1126/science.226.4681.1421}.

\bibitem[Lagrange et~al.(2009)Lagrange, Gratadour, Chauvin, Fusco, Ehrenreich, et~al.]{lagrange_probable_2009}
A.~M. Lagrange, D.~Gratadour, G.~Chauvin, T.~Fusco, D.~Ehrenreich, et~al.
\newblock A probable giant planet imaged in the β {Pictoris} disk. {VLT}/{NaCo} deep {L}'-band imaging.
\newblock \emph{Astronomy and Astrophysics}, 493:\penalty0 L21--L25, January 2009.
\newblock ISSN 0004-6361.
\newblock \doi{10.1051/0004-6361:200811325}.

\bibitem[Lagrange et~al.(2010)Lagrange, Bonnefoy, Chauvin, Apai, Ehrenreich, et~al.]{lagrange_giant_2010}
A.-M. Lagrange, M.~Bonnefoy, G.~Chauvin, D.~Apai, D.~Ehrenreich, et~al.
\newblock A {Giant} {Planet} {Imaged} in the {Disk} of the {Young} {Star} β {Pictoris}.
\newblock \emph{Science}, 329:\penalty0 57, July 2010.
\newblock ISSN 0036-8075.
\newblock \doi{10.1126/science.1187187}.

\bibitem[Lagrange et~al.(2019)Lagrange, Meunier, Rubini, Keppler, Galland, et~al.]{lagrange_evidence_2019}
A.~M. Lagrange, Nadège Meunier, Pascal Rubini, Miriam Keppler, Franck Galland, et~al.
\newblock Evidence for an additional planet in the β {Pictoris} system.
\newblock \emph{Nature Astronomy}, 3:\penalty0 1135--1142, August 2019.
\newblock ISSN 2397-3366.
\newblock \doi{10.1038/s41550-019-0857-1}.

\bibitem[Gibbs et~al.(2026)Gibbs, Ruffio, Bidot, Barman, Do~Ó, et~al.]{gibbs_discovery_2026}
Aidan Gibbs, Jean-Baptiste Ruffio, Alexis Bidot, Travis~S. Barman, Clarissa~R. Do~Ó, et~al.
\newblock Discovery of an {Exterior} {Third} {Planet} {Orbiting} β {Pictoris}.
\newblock \emph{The Astrophysical Journal Letters}, 1006:\penalty0 L11, July 2026.
\newblock ISSN 0004-637X.
\newblock \doi{10.3847/2041-8213/ae801b}.

\bibitem[Sutlieff et~al.(2026)Sutlieff, Bonse, Christiaens, Fontanive, Matthews, et~al.]{sutlieff_direct_2026}
Ben~J. Sutlieff, Markus~J. Bonse, Valentin Christiaens, Clémence Fontanive, Elisabeth~C. Matthews, et~al.
\newblock Direct {Imaging} {Discovery} of {Giant} {Exoplanet} β {Pictoris} d: {A} {Decade}-long {Game} of {Hide}-and-seek.
\newblock \emph{The Astrophysical Journal Letters}, 1006:\penalty0 L10, July 2026.
\newblock ISSN 0004-637X.
\newblock \doi{10.3847/2041-8213/ae80a0}.

\bibitem[Lacour et~al.(2021)Lacour, Wang, Rodet, Nowak, Shangguan, et~al.]{lacour_mass_2021}
S.~Lacour, J.~J. Wang, L.~Rodet, M.~Nowak, J.~Shangguan, et~al.
\newblock The mass of β {Pictoris} c from β {Pictoris} b orbital motion.
\newblock \emph{Astronomy and Astrophysics}, 654:\penalty0 L2, October 2021.
\newblock ISSN 0004-6361.
\newblock \doi{10.1051/0004-6361/202141889}.

\bibitem[Zwintz et~al.(2019)Zwintz, Reese, Neiner, Pigulski, Kuschnig, et~al.]{zwintz_revisiting_2019}
K.~Zwintz, D.~R. Reese, C.~Neiner, A.~Pigulski, R.~Kuschnig, et~al.
\newblock Revisiting the pulsational characteristics of the exoplanet host star β {Pictoris}.
\newblock \emph{Astronomy and Astrophysics}, 627:\penalty0 A28, July 2019.
\newblock ISSN 0004-6361.
\newblock \doi{10.1051/0004-6361/201834744}.

\bibitem[Chilcote et~al.(2017)Chilcote, Pueyo, De~Rosa, Vargas, Macintosh, et~al.]{chilcote_1-24_2017}
Jeffrey Chilcote, Laurent Pueyo, Robert~J. De~Rosa, Jeffrey Vargas, Bruce Macintosh, et~al.
\newblock 1-2.4 μm {Near}-{IR} {Spectrum} of the {Giant} {Planet} β {Pictoris} b {Obtained} with the {Gemini} {Planet} {Imager}.
\newblock \emph{The Astronomical Journal}, 153:\penalty0 182, April 2017.
\newblock ISSN 0004-6256.
\newblock \doi{10.3847/1538-3881/aa63e9}.

\bibitem[Zhang et~al.(2020)Zhang, Hallinan, Brisken, Bourke, and Golden]{zhang_multiepoch_2020}
Qicheng Zhang, Gregg Hallinan, Walter Brisken, Stephen Bourke, and Aaron Golden.
\newblock Multiepoch {VLBI} of {L} {Dwarf} {Binary} {2MASS} {J0746}+{2000AB}: {Precise} {Mass} {Measurements} and {Confirmation} of {Radio} {Emission} from {Both} {Components}.
\newblock \emph{The Astrophysical Journal}, 897:\penalty0 11, July 2020.
\newblock ISSN 0004-637X.
\newblock \doi{10.3847/1538-4357/ab9177}.

\bibitem[Driessen et~al.(2022)Driessen, Stappers, Tremou, Fender, Woudt, et~al.]{driessen_21_2022}
L.~N. Driessen, B.~W. Stappers, E.~Tremou, R.~P. Fender, P.~A. Woudt, et~al.
\newblock 21 new long-term variables in the {GX} 339-4 field: two years of {MeerKAT} monitoring.
\newblock \emph{Monthly Notices of the Royal Astronomical Society}, 512:\penalty0 5037--5066, June 2022.
\newblock ISSN 0035-8711.
\newblock \doi{10.1093/mnras/stac756}.

\bibitem[Driessen et~al.(2024{\natexlab{a}})Driessen, Barr, Buckley, Caleb, Chen, et~al.]{driessen_frb_2024}
L.~N. Driessen, E.~D. Barr, D.~A.~H. Buckley, M.~Caleb, H.~Chen, et~al.
\newblock {FRB} {20210405I}: a nearby {Fast} {Radio} {Burst} localized to sub-arcsecond precision with {MeerKAT}.
\newblock \emph{Monthly Notices of the Royal Astronomical Society}, 527:\penalty0 3659--3673, January 2024{\natexlab{a}}.
\newblock ISSN 0035-8711.
\newblock \doi{10.1093/mnras/stad3329}.

\bibitem[Dulk(1985)]{dulk_radio_1985}
G.~A. Dulk.
\newblock Radio emission from the sun and stars.
\newblock \emph{Annual Review of Astronomy and Astrophysics}, 23:\penalty0 169--224, January 1985.
\newblock ISSN 0066-4146.
\newblock \doi{10.1146/annurev.aa.23.090185.001125}.

\bibitem[Christensen et~al.(2009)Christensen, Holzwarth, and Reiners]{christensen_energy_2009}
Ulrich~R. Christensen, Volkmar Holzwarth, and Ansgar Reiners.
\newblock Energy flux determines magnetic field strength of planets and stars.
\newblock \emph{Nature}, 457:\penalty0 167--169, January 2009.
\newblock ISSN 0028-0836.
\newblock \doi{10.1038/nature07626}.

\bibitem[Günther et~al.(2012)Günther, Wolk, Drake, Lisse, Robrade, et~al.]{gunther_soft_2012}
H.~M. Günther, S.~J. Wolk, J.~J. Drake, C.~M. Lisse, J.~Robrade, et~al.
\newblock Soft {Coronal} {X}-{Rays} from β {Pictoris}.
\newblock \emph{The Astrophysical Journal}, 750:\penalty0 78, May 2012.
\newblock ISSN 0004-637X.
\newblock \doi{10.1088/0004-637X/750/1/78}.

\bibitem[Williams et~al.(2014)Williams, Cook, and Berger]{williams_trends_2014}
P.~K.~G. Williams, B.~A. Cook, and E.~Berger.
\newblock Trends in {Ultracool} {Dwarf} {Magnetism}. {I}. {X}-{Ray} {Suppression} and {Radio} {Enhancement}.
\newblock \emph{The Astrophysical Journal}, 785:\penalty0 9, April 2014.
\newblock ISSN 0004-637X.
\newblock \doi{10.1088/0004-637X/785/1/9}.

\bibitem[Magaudda et~al.(2024)Magaudda, Stelzer, Osten, Pineda, Raetz, et~al.]{magaudda_transitions_2024}
E.~Magaudda, B.~Stelzer, R.~A. Osten, J.~S. Pineda, St. Raetz, et~al.
\newblock Transitions in magnetic behavior at the substellar boundary.
\newblock \emph{Astronomy and Astrophysics}, 687:\penalty0 A95, July 2024.
\newblock ISSN 0004-6361.
\newblock \doi{10.1051/0004-6361/202449403}.

\bibitem[Hill(1979)]{hill_inertial_1979}
T.~W. Hill.
\newblock Inertial limit on corotation.
\newblock \emph{Journal of Geophysical Research}, 84:\penalty0 6554--6558, November 1979.
\newblock ISSN 0148-0227.
\newblock \doi{10.1029/JA084iA11p06554}.

\bibitem[Cowley and Bunce(2001)]{cowley_origin_2001}
S.~W.~H. Cowley and E.~J. Bunce.
\newblock Origin of the main auroral oval in {Jupiter}'s coupled magnetosphere-ionosphere system.
\newblock \emph{Planetary and Space Science}, 49:\penalty0 1067--1088, August 2001.
\newblock ISSN 0032-0633.
\newblock \doi{10.1016/S0032-0633(00)00167-7}.

\bibitem[Nichols et~al.(2012)Nichols, Burleigh, Casewell, Cowley, Wynn, et~al.]{nichols_origin_2012}
J.~D. Nichols, M.~R. Burleigh, S.~L. Casewell, S.~W.~H. Cowley, G.~A. Wynn, et~al.
\newblock Origin of {Electron} {Cyclotron} {Maser} {Induced} {Radio} {Emissions} at {Ultracool} {Dwarfs}: {Magnetosphere}-{Ionosphere} {Coupling} {Currents}.
\newblock \emph{The Astrophysical Journal}, 760:\penalty0 59, November 2012.
\newblock ISSN 0004-637X.
\newblock \doi{10.1088/0004-637X/760/1/59}.

\bibitem[Snellen et~al.(2014)Snellen, Brandl, de~Kok, Brogi, Birkby, et~al.]{snellen_fast_2014}
Ignas A.~G. Snellen, Bernhard~R. Brandl, Remco~J. de~Kok, Matteo Brogi, Jayne Birkby, et~al.
\newblock Fast spin of the young extrasolar planet β {Pictoris} b.
\newblock \emph{Nature}, 509:\penalty0 63--65, May 2014.
\newblock ISSN 0028-0836.
\newblock \doi{10.1038/nature13253}.

\bibitem[Landman et~al.(2024)Landman, Stolker, Snellen, Costes, de~Regt, et~al.]{landman__2024}
R.~Landman, T.~Stolker, I.~A.~G. Snellen, J.~Costes, S.~de~Regt, et~al.
\newblock β {Pictoris} b through the eyes of the upgraded {CRIRES}+. {Atmospheric} composition, spin rotation, and radial velocity.
\newblock \emph{Astronomy and Astrophysics}, 682:\penalty0 A48, February 2024.
\newblock ISSN 0004-6361.
\newblock \doi{10.1051/0004-6361/202347846}.

\bibitem[Janson et~al.(2025)Janson, Wehrung-Montpezat, Wehrhahn, Brandeker, Viswanath, et~al.]{janson_deep_2025}
Markus Janson, Jonas Wehrung-Montpezat, Ansgar Wehrhahn, Alexis Brandeker, Gayathri Viswanath, et~al.
\newblock Deep high-resolution {L} band spectroscopy in the β {Pictoris} planetary system.
\newblock \emph{Astronomy and Astrophysics}, 694:\penalty0 A63, February 2025.
\newblock ISSN 0004-6361.
\newblock \doi{10.1051/0004-6361/202452411}.

\bibitem[Zhou et~al.(2026)Zhou, Biller, Carter, Perrin, Poon, et~al.]{zhou_photometric_2026}
Yifan Zhou, Beth~A. Biller, Aarynn~L. Carter, Marshall~D. Perrin, Michael Poon, et~al.
\newblock Photometric {Variability} and {Rotation} of {Beta} {Pictoris} b from {JWST} {NIRCam} {Coronagraphic} {Imaging}.
\newblock arXiv:2607.13133, July 2026.

\bibitem[Barrow and Desch(1989)]{barrow_solar_1989}
C.~H. Barrow and M.~D. Desch.
\newblock Solar wind control of {Jupiter}'s hectometric radio emission.
\newblock \emph{Astronomy and Astrophysics}, 213:\penalty0 495--501, April 1989.
\newblock ISSN 0004-6361.

\bibitem[Katarzyński et~al.(2016)Katarzyński, Gawroński, and Goździewski]{katarzynski_search_2016}
K.~Katarzyński, M.~Gawroński, and K.~Goździewski.
\newblock Search for exoplanets and brown dwarfs with {VLBI}.
\newblock \emph{Monthly Notices of the Royal Astronomical Society}, 461:\penalty0 929--938, September 2016.
\newblock ISSN 0035-8711.
\newblock \doi{10.1093/mnras/stw1354}.

\bibitem[Ashtari et~al.(2022)Ashtari, Sciola, Turner, and Stevenson]{ashtari_detecting_2022}
Reza Ashtari, Anthony Sciola, Jake~D. Turner, and Kevin Stevenson.
\newblock Detecting {Magnetospheric} {Radio} {Emission} from {Giant} {Exoplanets}.
\newblock \emph{The Astrophysical Journal}, 939:\penalty0 24, November 2022.
\newblock ISSN 0004-637X.
\newblock \doi{10.3847/1538-4357/ac92f5}.

\bibitem[Grießmeier et~al.(2005)Grießmeier, Motschmann, Mann, and Rucker]{griesmeier_influence_2005}
J.~M. Grießmeier, U.~Motschmann, G.~Mann, and H.~O. Rucker.
\newblock The influence of stellar wind conditions on the detectability of planetary radio emissions.
\newblock \emph{Astronomy and Astrophysics}, 437:\penalty0 717--726, July 2005.
\newblock ISSN 0004-6361.
\newblock \doi{10.1051/0004-6361:20041976}.

\bibitem[Bruhweiler et~al.(1991)Bruhweiler, Kondo, and Grady]{bruhweiler_mass_1991}
Frederick~C. Bruhweiler, Yoji Kondo, and C.~A. Grady.
\newblock Mass {Outflow} in the {Nearby} {Proto}--{Planetary} {System} beta {Pictoris}.
\newblock \emph{The Astrophysical Journal}, 371:\penalty0 L27, April 1991.
\newblock ISSN 0004-637X.
\newblock \doi{10.1086/185994}.

\bibitem[Bigg(1964)]{bigg_influence_1964}
E.~K. Bigg.
\newblock Influence of the {Satellite} {Io} on {Jupiter}'s {Decametric} {Emission}.
\newblock \emph{Nature}, 203:\penalty0 1008--1010, September 1964.
\newblock ISSN 0028-0836.
\newblock \doi{10.1038/2031008a0}.

\bibitem[Goldreich and Lynden-Bell(1969)]{goldreich_io_1969}
P.~Goldreich and D.~Lynden-Bell.
\newblock Io, a jovian unipolar inductor.
\newblock \emph{The Astrophysical Journal}, 156:\penalty0 59--78, April 1969.
\newblock ISSN 0004-637X.
\newblock \doi{10.1086/149947}.

\bibitem[Macias et~al.(2026)Macias, Jenkins, and Vanderburg]{macias_first_2026}
Isabella Macias, Sydney~A. Jenkins, and Andrew Vanderburg.
\newblock First {Astrometric} {Limits} on {Binary} {Planets} and {Exomoons} {Orbiting} β {Pictoris} b.
\newblock \emph{The Astronomical Journal}, 171:\penalty0 197, March 2026.
\newblock ISSN 0004-6256.
\newblock \doi{10.3847/1538-3881/ae421c}.

\bibitem[Kenworthy et~al.(2026)Kenworthy, Landman, Vanderburg, Rodriguez, Birkby, et~al.]{kenworthy_upper_2026}
Matthew~A. Kenworthy, Rico Landman, Andrew Vanderburg, Joseph~E. Rodriguez, Jayne~L. Birkby, et~al.
\newblock Upper limits on exosatellites around β {Pictoris} b.
\newblock \emph{Monthly Notices of the Royal Astronomical Society}, 549:\penalty0 stag1060, July 2026.
\newblock ISSN 0035-8711.
\newblock \doi{10.1093/mnras/stag1060}.

\bibitem[Reiners and Christensen(2010)]{reiners_magnetic_2010}
A.~Reiners and U.~R. Christensen.
\newblock A magnetic field evolution scenario for brown dwarfs and giant planets.
\newblock \emph{Astronomy \& Astrophysics}, 522:\penalty0 A13, October 2010.
\newblock ISSN 0004-6361.
\newblock \doi{10.1051/0004-6361/201014251}.

\bibitem[Kavanagh et~al.(2024)Kavanagh, Vedantham, Rose, and Bloot]{kavanagh_unravelling_2024}
Robert~D. Kavanagh, Harish~K. Vedantham, Kovi Rose, and Sanne Bloot.
\newblock Unravelling sub-stellar magnetospheres.
\newblock \emph{Astronomy and Astrophysics}, 692:\penalty0 A66, December 2024.
\newblock ISSN 0004-6361.
\newblock \doi{10.1051/0004-6361/202452094}.

\bibitem[Offringa et~al.(2014)Offringa, McKinley, Hurley-Walker, Briggs, Wayth, et~al.]{offringa_wsclean_2014}
A.~R. Offringa, B.~McKinley, N.~Hurley-Walker, F.~H. Briggs, R.~B. Wayth, et~al.
\newblock {WSCLEAN}: an implementation of a fast, generic wide-field imager for radio astronomy.
\newblock \emph{Monthly Notices of the Royal Astronomical Society}, 444:\penalty0 606--619, October 2014.
\newblock ISSN 0035-8711.
\newblock \doi{10.1093/mnras/stu1368}.

\bibitem[Offringa and Smirnov(2017)]{offringa_optimized_2017}
A.~R. Offringa and O.~Smirnov.
\newblock An optimized algorithm for multiscale wideband deconvolution of radio astronomical images.
\newblock \emph{Monthly Notices of the Royal Astronomical Society}, 471:\penalty0 301--316, October 2017.
\newblock ISSN 0035-8711.
\newblock \doi{10.1093/mnras/stx1547}.

\bibitem[Ranchod et~al.(2025)Ranchod, Wagenveld, Klöckner, Wucknitz, Deane, et~al.]{ranchod_first_2025}
S.~Ranchod, J.~D. Wagenveld, H.-R. Klöckner, O.~Wucknitz, R.~P. Deane, et~al.
\newblock A first glimpse at the {MeerKAT} {DEEP2} field at {S}-band.
\newblock \emph{Monthly Notices of the Royal Astronomical Society}, 536:\penalty0 3647--3662, February 2025.
\newblock ISSN 0035-8711.
\newblock \doi{10.1093/mnras/stae2754}.

\bibitem[Hughes et~al.(2025{\natexlab{a}})Hughes, Carotenuto, Russell, Tetarenko, Miller-Jones, et~al.]{hughes_comprehensive_2025}
Andrew~K. Hughes, Francesco Carotenuto, Thomas~D. Russell, Alexandra~J. Tetarenko, James C.~A. Miller-Jones, et~al.
\newblock Comprehensive {Radio} {Monitoring} of the {Black} {Hole} {X}-{Ray} {Binary} {Swift} {J1727}.8−1613 during {Its} 2023─2024 {Outburst}.
\newblock \emph{The Astrophysical Journal}, 988:\penalty0 109, July 2025{\natexlab{a}}.
\newblock ISSN 0004-637X.
\newblock \doi{10.3847/1538-4357/ade2e6}.

\bibitem[Hughes et~al.(2025{\natexlab{b}})Hughes, Cowie, Heywood, and Hugo]{hughes_polkat_2025}
Andrew~K. Hughes, Fraser~J. Cowie, Ian Heywood, and Ben Hugo.
\newblock polkat: {Semi}-automate full polarization of {MeerKAT} observations.
\newblock Astrophysics Source Code Library, \texttt{ascl:2502.026}, February 2025{\natexlab{b}}.

\bibitem[Heywood(2020)]{heywood_oxkat_2020}
Ian Heywood.
\newblock oxkat: {Semi}-automated imaging of {MeerKAT} observations.
\newblock Astrophysics Source Code Library, \texttt{ascl:2009.003}, September 2020.

\bibitem[{CASA Team} et~al.(2022){CASA Team}, Bean, Bhatnagar, Castro, Donovan~Meyer, et~al.]{casa_team_casa_2022}
{CASA Team}, Ben Bean, Sanjay Bhatnagar, Sandra Castro, Jennifer Donovan~Meyer, et~al.
\newblock {CASA}, the {Common} {Astronomy} {Software} {Applications} for {Radio} {Astronomy}.
\newblock \emph{Publications of the Astronomical Society of the Pacific}, 134:\penalty0 114501, November 2022.
\newblock ISSN 0004-6280.
\newblock \doi{10.1088/1538-3873/ac9642}.

\bibitem[Hugo et~al.(2022)Hugo, Perkins, Merry, Mauch, and Smirnov]{hugo_tricolour_2022}
Benjamin~V. Hugo, S.~Perkins, B.~Merry, T.~Mauch, and O.~M. Smirnov.
\newblock Tricolour: {An} {Optimized} {SumThreshold} {Flagger} for {MeerKAT}.
\newblock arXiv:2206.09179, July 2022.

\bibitem[Kenyon et~al.(2025)Kenyon, Perkins, Bester, Smirnov, Russeeawon, et~al.]{kenyon_africanus_2025}
J.~S. Kenyon, S.~J. Perkins, H.~L. Bester, O.~M. Smirnov, C.~Russeeawon, et~al.
\newblock Africanus {II}. {QuartiCal}: {Calibrating} radio interferometer data at scale using {Numba} and {Dask}.
\newblock \emph{Astronomy and Computing}, 52:\penalty0 100962, July 2025.
\newblock ISSN 2213-1337.
\newblock \doi{10.1016/j.ascom.2025.100962}.

\bibitem[Hales(2017)]{hales_calibration_2017}
Christopher~A. Hales.
\newblock Calibration {Errors} in {Interferometric} {Radio} {Polarimetry}.
\newblock \emph{The Astronomical Journal}, 154:\penalty0 54, August 2017.
\newblock ISSN 0004-6256.
\newblock \doi{10.3847/1538-3881/aa7aef}.

\bibitem[Wang et~al.(2021)Wang, Kulikauskas, and Blunt]{wang_whereistheplanet_2021}
Jason~J. Wang, Matas Kulikauskas, and Sarah Blunt.
\newblock whereistheplanet: {Predicting} positions of directly imaged companions.
\newblock \emph{Astrophysics Source Code Library}, page ascl:2101.003, January 2021.

\bibitem[Hancock et~al.(2012)Hancock, Murphy, Gaensler, Hopkins, and Curran]{hancock_compact_2012}
P.~J. Hancock, T.~Murphy, B.~M. Gaensler, A.~Hopkins, and J.~R. Curran.
\newblock Compact continuum source finding for next generation radio surveys.
\newblock \emph{Monthly Notices of the Royal Astronomical Society}, 422:\penalty0 1812--1824, May 2012.
\newblock ISSN 0035-8711.
\newblock \doi{10.1111/j.1365-2966.2012.20768.x}.

\bibitem[Hancock et~al.(2018)Hancock, Trott, and Hurley-Walker]{hancock_source_2018}
Paul~J. Hancock, Cathryn~M. Trott, and Natasha Hurley-Walker.
\newblock Source {Finding} in the {Era} of the {SKA} ({Precursors}): {Aegean} 2.0.
\newblock \emph{Publications of the Astronomical Society of Australia}, 35:\penalty0 e011, March 2018.
\newblock ISSN 1323-3580.
\newblock \doi{10.1017/pasa.2018.3}.

\bibitem[Charlot et~al.(2020)Charlot, Jacobs, Gordon, Lambert, de~Witt, et~al.]{charlot_third_2020}
P.~Charlot, C.~S. Jacobs, D.~Gordon, S.~Lambert, A.~de~Witt, et~al.
\newblock The third realization of the {International} {Celestial} {Reference} {Frame} by very long baseline interferometry.
\newblock \emph{Astronomy and Astrophysics}, 644:\penalty0 A159, December 2020.
\newblock ISSN 0004-6361.
\newblock \doi{10.1051/0004-6361/202038368}.

\bibitem[Petrov and Kovalev(2025)]{petrov_radio_2025}
L.~Y. Petrov and Y.~Y. Kovalev.
\newblock The {Radio} {Fundamental} {Catalog}. {I}. {Astrometry}.
\newblock \emph{The Astrophysical Journal Supplement Series}, 276:\penalty0 38, February 2025.
\newblock ISSN 0067-0049.
\newblock \doi{10.3847/1538-4365/ad8c36}.

\bibitem[{Gaia Collaboration} et~al.(2022){Gaia Collaboration}, Klioner, Lindegren, Mignard, Hernández, et~al.]{gaia_collaboration_gaia_2022}
{Gaia Collaboration}, S.~A. Klioner, L.~Lindegren, F.~Mignard, J.~Hernández, et~al.
\newblock Gaia {Early} {Data} {Release} 3. {The} celestial reference frame ({Gaia}-{CRF3}).
\newblock \emph{Astronomy and Astrophysics}, 667:\penalty0 A148, November 2022.
\newblock ISSN 0004-6361.
\newblock \doi{10.1051/0004-6361/202243483}.

\bibitem[Petrov et~al.(2019)Petrov, de~Witt, Sadler, Phillips, and Horiuchi]{petrov_second_2019}
Leonid Petrov, Alet de~Witt, Elaine~M. Sadler, Chris Phillips, and Shinji Horiuchi.
\newblock The {Second} {LBA} {Calibrator} {Survey} of southern compact extragalactic radio sources - {LCS2}.
\newblock \emph{Monthly Notices of the Royal Astronomical Society}, 485:\penalty0 88--101, May 2019.
\newblock ISSN 0035-8711.
\newblock \doi{10.1093/mnras/stz242}.

\bibitem[Condon(1997)]{condon_errors_1997}
J.~J. Condon.
\newblock Errors in {Elliptical} {Gaussian} {Fits}.
\newblock \emph{Publications of the Astronomical Society of the Pacific}, 109:\penalty0 166--172, February 1997.
\newblock ISSN 0004-6280.
\newblock \doi{10.1086/133871}.

\bibitem[Loi et~al.(2015)Loi, Murphy, Bell, Kaplan, Lenc, et~al.]{loi_quantifying_2015}
Shyeh~Tjing Loi, Tara Murphy, Martin~E. Bell, David~L. Kaplan, Emil Lenc, et~al.
\newblock Quantifying ionospheric effects on time-domain astrophysics with the {Murchison} {Widefield} {Array}.
\newblock \emph{Monthly Notices of the Royal Astronomical Society}, 453:\penalty0 2731--2746, November 2015.
\newblock ISSN 0035-8711.
\newblock \doi{10.1093/mnras/stv1808}.

\bibitem[Cohen and Röttgering(2009)]{cohen_probing_2009}
A.~S. Cohen and H.~J.~A. Röttgering.
\newblock Probing {Fine}-{Scale} {Ionospheric} {Structure} with the {Very} {Large} {Array} {Radio} {Telescope}.
\newblock \emph{The Astronomical Journal}, 138:\penalty0 439--447, August 2009.
\newblock ISSN 0004-6256.
\newblock \doi{10.1088/0004-6256/138/2/439}.

\bibitem[Morzinski et~al.(2015)Morzinski, Males, Skemer, Close, Hinz, et~al.]{morzinski_magellan_2015}
Katie~M. Morzinski, Jared~R. Males, Andy~J. Skemer, Laird~M. Close, Phil~M. Hinz, et~al.
\newblock Magellan {Adaptive} {Optics} {First}-light {Observations} of the {Exoplanet} β {Pic} b. {II}. 3-5 μm {Direct} {Imaging} with {MagAO}+{Clio}, and the {Empirical} {Bolometric} {Luminosity} of a {Self}-luminous {Giant} {Planet}.
\newblock \emph{The Astrophysical Journal}, 815:\penalty0 108, December 2015.
\newblock ISSN 0004-637X.
\newblock \doi{10.1088/0004-637X/815/2/108}.

\bibitem[Hallinan et~al.(2008)Hallinan, Antonova, Doyle, Bourke, Lane, et~al.]{hallinan_confirmation_2008}
G.~Hallinan, A.~Antonova, J.~G. Doyle, S.~Bourke, C.~Lane, et~al.
\newblock Confirmation of the {Electron} {Cyclotron} {Maser} {Instability} as the {Dominant} {Source} of {Radio} {Emission} from {Very} {Low} {Mass} {Stars} and {Brown} {Dwarfs}.
\newblock \emph{The Astrophysical Journal}, 684:\penalty0 644--653, September 2008.
\newblock ISSN 0004-637X.
\newblock \doi{10.1086/590360}.

\bibitem[Connerney et~al.(2022)Connerney, Timmins, Oliversen, Espley, Joergensen, et~al.]{connerney_new_2022}
J.~E.~P. Connerney, S.~Timmins, R.~J. Oliversen, J.~R. Espley, J.~L. Joergensen, et~al.
\newblock A {New} {Model} of {Jupiter}'s {Magnetic} {Field} at the {Completion} of {Juno}'s {Prime} {Mission}.
\newblock \emph{Journal of Geophysical Research (Planets)}, 127:\penalty0 e07055, February 2022.
\newblock ISSN 0148-0227.
\newblock \doi{10.1029/2021JE007055}.

\bibitem[Ricci et~al.(2015)Ricci, Maddison, Wilner, MacGregor, Ubach, et~al.]{ricci_atca_2015}
L.~Ricci, S.~T. Maddison, D.~Wilner, M.~A. MacGregor, C.~Ubach, et~al.
\newblock An {ATCA} {Survey} of {Debris} {Disks} at 7 {Millimeters}.
\newblock \emph{The Astrophysical Journal}, 813:\penalty0 138, November 2015.
\newblock ISSN 0004-637X.
\newblock \doi{10.1088/0004-637X/813/2/138}.

\bibitem[Leto et~al.(2021)Leto, Trigilio, Krtička, Fossati, Ignace, et~al.]{leto_scaling_2021}
P.~Leto, C.~Trigilio, J.~Krtička, L.~Fossati, R.~Ignace, et~al.
\newblock A scaling relationship for non-thermal radio emission from ordered magnetospheres: from the top of the main sequence to planets.
\newblock \emph{Monthly Notices of the Royal Astronomical Society}, 507:\penalty0 1979--1998, October 2021.
\newblock ISSN 0035-8711.
\newblock \doi{10.1093/mnras/stab2168}.

\bibitem[Saffe et~al.(2021)Saffe, Miquelarena, Alacoria, Flores, Jaque~Arancibia, et~al.]{saffe_chemical_2021}
C.~Saffe, P.~Miquelarena, J.~Alacoria, M.~Flores, M.~Jaque~Arancibia, et~al.
\newblock Chemical analysis of early-type stars with planets.
\newblock \emph{Astronomy and Astrophysics}, 647:\penalty0 A49, March 2021.
\newblock ISSN 0004-6361.
\newblock \doi{10.1051/0004-6361/202040132}.

\bibitem[White et~al.(2021)White, Tapia-Vázquez, Hughes, Moór, Matthews, et~al.]{white_first_2021}
Jacob~Aaron White, F.~Tapia-Vázquez, A.~G. Hughes, A.~Moór, B.~Matthews, et~al.
\newblock The {First} {Radio} {Spectrum} of a {Rapidly} {Rotating} {A}-type {Star}.
\newblock \emph{The Astrophysical Journal}, 912:\penalty0 L5, May 2021.
\newblock ISSN 0004-637X.
\newblock \doi{10.3847/2041-8213/abf6da}.

\bibitem[Owocki et~al.(2022)Owocki, Shultz, ud~Doula, Chandra, Das, et~al.]{owocki_centrifugal_2022}
S.~P. Owocki, M.~E. Shultz, A.~ud~Doula, P.~Chandra, B.~Das, et~al.
\newblock Centrifugal breakout reconnection as the electron acceleration mechanism powering the radio magnetospheres of early-type stars.
\newblock \emph{Monthly Notices of the Royal Astronomical Society}, 513:\penalty0 1449--1458, June 2022.
\newblock ISSN 0035-8711.
\newblock \doi{10.1093/mnras/stac341}.

\bibitem[Das et~al.(2022)Das, Chandra, Shultz, Wade, Sikora, et~al.]{das_discovery_2022}
Barnali Das, Poonam Chandra, Matt~E. Shultz, Gregg~A. Wade, James Sikora, et~al.
\newblock Discovery of {Eight} "{Main}-sequence {Radio} {Pulse} {Emitters}" {Using} the {GMRT}: {Clues} to the {Onset} of {Coherent} {Radio} {Emission} in {Hot} {Magnetic} {Stars}.
\newblock \emph{The Astrophysical Journal}, 925:\penalty0 125, February 2022.
\newblock ISSN 0004-637X.
\newblock \doi{10.3847/1538-4357/ac2576}.

\bibitem[Guedel and Benz(1993)]{guedel_x-raymicrowave_1993}
Manuel Guedel and Arnold~O. Benz.
\newblock X-{Ray}/{Microwave} {Relation} of {Different} {Types} of {Active} {Stars}.
\newblock \emph{The Astrophysical Journal Letters}, 405:\penalty0 L63, March 1993.
\newblock ISSN 0004-637X.
\newblock \doi{10.1086/186766}.

\bibitem[Benz and Guedel(1994)]{benz_x-raymicrowave_1994}
A.~O. Benz and M.~Guedel.
\newblock X-ray/microwave ratio of flares and coronae.
\newblock \emph{Astronomy and Astrophysics}, 285:\penalty0 621--630, May 1994.
\newblock ISSN 0004-6361.

\bibitem[Hempel et~al.(2005)Hempel, Robrade, Ness, and Schmitt]{hempel_detection_2005}
M.~Hempel, J.~Robrade, J.~U. Ness, and J.~H. M.~M. Schmitt.
\newblock Detection of {X}-ray emission from β {Pictoris} with {XMM}-{Newton}: a cool corona, a boundary layer or what?
\newblock \emph{Astronomy and Astrophysics}, 440:\penalty0 727--734, September 2005.
\newblock ISSN 0004-6361.
\newblock \doi{10.1051/0004-6361:20042596}.

\bibitem[Schröder and Schmitt(2007)]{schroder_x-ray_2007}
C.~Schröder and J.~H. M.~M. Schmitt.
\newblock X-ray emission from {A}-type stars.
\newblock \emph{Astronomy and Astrophysics}, 475:\penalty0 677--684, November 2007.
\newblock ISSN 0004-6361.
\newblock \doi{10.1051/0004-6361:20077429}.

\bibitem[Günther et~al.(2022)Günther, Melis, Robrade, Schneider, Wolk, et~al.]{gunther_coronal_2022}
Hans~Moritz Günther, Carl Melis, J.~Robrade, P.~C. Schneider, Scott~J. Wolk, et~al.
\newblock Coronal and {Chromospheric} {Emission} in {A}-type {Stars}.
\newblock \emph{The Astronomical Journal}, 164:\penalty0 8, July 2022.
\newblock ISSN 0004-6256.
\newblock \doi{10.3847/1538-3881/ac6ef6}.

\bibitem[Gray et~al.(2006)Gray, Corbally, Garrison, McFadden, Bubar, et~al.]{gray_contributions_2006}
R.~O. Gray, C.~J. Corbally, R.~F. Garrison, M.~T. McFadden, E.~J. Bubar, et~al.
\newblock Contributions to the {Nearby} {Stars} ({NStars}) {Project}: {Spectroscopy} of {Stars} {Earlier} than {M0} within 40 pc-{The} {Southern} {Sample}.
\newblock \emph{The Astronomical Journal}, 132:\penalty0 161--170, July 2006.
\newblock ISSN 0004-6256.
\newblock \doi{10.1086/504637}.

\bibitem[{GRAVITY Collaboration} et~al.(2020){GRAVITY Collaboration}, Nowak, Lacour, Mollière, Wang, et~al.]{gravity_collaboration_peering_2020}
{GRAVITY Collaboration}, M.~Nowak, S.~Lacour, P.~Mollière, J.~Wang, et~al.
\newblock Peering into the formation history of β {Pictoris} b with {VLTI}/{GRAVITY} long-baseline interferometry.
\newblock \emph{Astronomy and Astrophysics}, 633:\penalty0 A110, January 2020.
\newblock ISSN 0004-6361.
\newblock \doi{10.1051/0004-6361/201936898}.

\bibitem[Nowak et~al.(2020)Nowak, Lacour, Lagrange, Rubini, Wang, et~al.]{nowak_direct_2020}
M.~Nowak, S.~Lacour, A.~M. Lagrange, P.~Rubini, J.~Wang, et~al.
\newblock Direct confirmation of the radial-velocity planet β {Pictoris} c.
\newblock \emph{Astronomy and Astrophysics}, 642:\penalty0 L2, October 2020.
\newblock ISSN 0004-6361.
\newblock \doi{10.1051/0004-6361/202039039}.

\bibitem[Driessen et~al.(2024{\natexlab{b}})Driessen, Pritchard, Murphy, Heald, Robrade, et~al.]{driessen_sydney_2024}
Laura~Nicole Driessen, Joshua Pritchard, Tara Murphy, George Heald, Jan Robrade, et~al.
\newblock The {Sydney} {Radio} {Star} {Catalogue}: {Properties} of radio stars at megahertz to gigahertz frequencies.
\newblock \emph{Publications of the Astronomical Society of Australia}, 41:\penalty0 e084, November 2024{\natexlab{b}}.
\newblock ISSN 1323-3580.
\newblock \doi{10.1017/pasa.2024.72}.

\bibitem[Freund et~al.(2022)Freund, Czesla, Robrade, Schneider, and Schmitt]{freund_stellar_2022}
S.~Freund, S.~Czesla, J.~Robrade, P.~C. Schneider, and J.~H. M.~M. Schmitt.
\newblock The stellar content of the {ROSAT} all-sky survey.
\newblock \emph{Astronomy and Astrophysics}, 664:\penalty0 A105, August 2022.
\newblock ISSN 0004-6361.
\newblock \doi{10.1051/0004-6361/202142573}.

\end{thebibliography}

\section*{Acknowledgements}

We thank S.~Andrews, B.~Das, L.~Driessen, and J.~Callingham for useful comments.
The MeerKAT telescope is operated by the South African Radio Astronomy Observatory, which is a facility of the National Research Foundation, an agency of the Department of Science, Technology and Innovation.
The Berger Time-Domain Group at Harvard is supported by NSF and NASA grants.
K.N.O.C.\ acknowledges the NSF Graduate Research Fellowship (DGE1745303) and the Ford Foundation Predoctoral Fellowship.
The computations in this paper were run on the FASRC Cannon cluster supported by the FAS Division of Science Research Computing Group at Harvard University. 
This work has made use of data from the European Space Agency (ESA) mission Gaia (\url{https://www.cosmos.esa.int/gaia}), processed by the Gaia Data Processing and Analysis Consortium (DPAC, \url{https://www.cosmos.esa.int/web/gaia/dpac/consortium}). Funding for the DPAC has been provided by national institutions, in particular the institutions participating in the Gaia Multilateral Agreement.
This work has made use of the ``MPIfR S-band receiver system'' designed, constructed and maintained by funding of the MPI für Radioastronomy and the Max-Planck-Society.

\section*{Competing interests}
The authors declare no competing interests.

\section*{Additional information}
Correspondence and requests for materials should be addressed to K.N.O.C.

\end{document}